\pdfoutput=1
\documentclass[useAMS,usenatbib]{mn2e}

\usepackage{graphicx}
\usepackage{amssymb}
\usepackage{multirow}

\makeatletter
    
    \newcommand{\Rmnum}[1]{\expandafter\@slowromancap\romannumeral #1@}
\makeatother

\title[A study of the high-inclination population in the Kuiper belt -- \Rmnum1. The Plutinos]
         {A study of the high-inclination population in the Kuiper belt -- \Rmnum1. The Plutinos}
\author[Jian Li, Li-Yong Zhou and Yi-Sui Sun]
{Jian Li\thanks{E-mail: ljian@nju.edu.cn},  Li-Yong Zhou and Yi-Sui Sun\\
School of Astronomy and Space Science
\& Key Laboratory of Modern Astronomy and Astrophysics in Ministry of Education,\\
 Nanjing University, Nanjing 210093, PR China}
\begin{document}

\date{Accepted 2013 October 2.  Received 2013 October 2; in original form 2012 December 10}

\maketitle

\label{firstpage}

\begin{abstract}

The dynamics of the high-inclination Plutinos is systematically studied. We first present the peculiar features of the 2:3 Neptune mean motion resonance (NMMR) for inclined orbits, especially for the correlation of resonant amplitude $A_{\sigma}$ with inclination $i$. Using the numerical integrations for the age of the Solar system, the dynamical structure of the 2:3 NMMR is mapped out on the plane of semi-major axis versus $i$ for different eccentricities. We have shown that $i$ of stable resonant orbits could be as high as $90^{\circ}$; and the stable region is roughly surrounded by the contours of $A_{\sigma}=120^{\circ}$. These new findings allow us to further explore the 2:3 NMMR capture and retention of planetesimals with initial inclinations $i_0 \le 90^{\circ}$ in the frame of the planet migration model. We find that the outward transportation of Plutinos is possible for any inclined or even perpendicular orbits. 
 
The role of $i_0$ in the formation of Plutinos during Neptune's migration is highlighted and interesting results are obtained: (1) The capture efficiency of the 2:3 NMMR decreases drastically first with the increase of $i_0$, but it then raises instead when $i_0$ exceeds $\sim50^{\circ}$; (2) The magnitude of $i$-variation is limited to less than $5^{\circ}$ for any $i_0$, and moreover, for Plutinos with $i\gtrsim48^{\circ}$, their $i$ are forced to decrease throughout the outward migration; (3) Plutinos with $i\gtrsim48^{\circ}$ are certainly outside the Kozai mechanism, since an inclination increase is prohibited by the migrating 2:3 NMMR; (4) The 7:11 inclination-type NMMR could be responsible for nearly-circular Plutinos, and a minimum $i_0\sim15^{\circ}$ is required to intrigue this mechanism.  

 \end{abstract}



\section{Introduction}

\begin{figure}
 \hspace{-0.7cm}
  \centering
  \includegraphics[width=9cm]{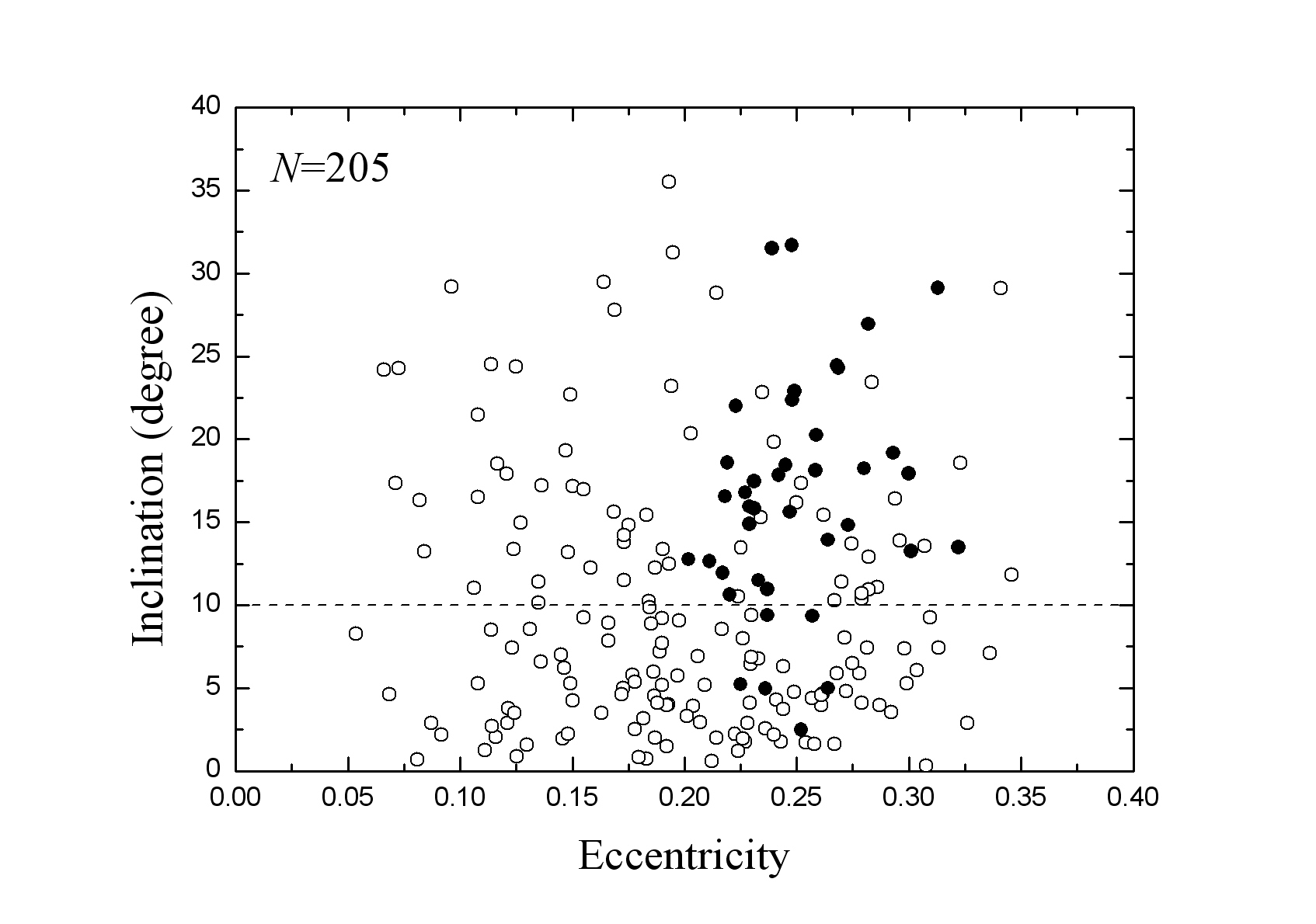}
  \caption{Distribution of eccentricities and inclinations for the currently observed Plutinos (as of October 2012). The filled circles stand for Plutinos experiencing the Kozai mechanism. Above the dashed line at $10^{\circ}$, Plutinos with high inclinations are most concerned in this paper. }
  \label{observed}
\end{figure}

The Plutinos are a set of Kuiper belt objects (KBOs) who reside in the 2:3 Neptune mean motion resonance (NMMR) at semimajor axis $a\approx39.4$\,AU. Their dynamical properties are valuable clues as to what happened during the formation of the outer Solar system, thus have received the most interest in the study of the Kuiper belt (Malhotra et al. 2000; Luu \& Jewitt 2002; Morbidelli \& Brown 2004; Gladman 2005; Gomes 2009; Petit et al. 2011; Gladman et al. 2012). A generic feature of Plutinos is that a substantial number of them have perihelia inside the orbit of Neptune, while they are phase protected from close encounters with Neptune, e.g., the distance between Pluto and Neptune is never less than 18\,AU (Cohen \& Hubbard 1965). At the time of writing, more than 200 Plutinos have been discovered, and their eccentricities and inclinations at epoch 2012 September 30 are shown in Fig. \ref{observed}. To well identify these Plutinos, we take into account the KBOs with 39\,AU$\le a \le$40\,AU registered in the IAU Minor Planet Center\footnote{http://www.minorplanetcenter.net/iau/lists/TNOs.html}, and their orbits are integrated numerically by considering the gravitational effects of four Jovian planets. Then an object is regarded as Plutino if its resonant angle librates throughout the 10 Myr evolution. The mechanism of resonant capture envisioned by Malhotra (1995) was first proposed to explain the concentration of KBOs in the 2:3 NMMR. As Neptune migrated radially outwards due to the exchange of energy and angular momentum between the Jovian planets and planetesimals, its mean motion resonances swept through the residual disk and a lot of small bodies were captured. 

Yet, the planet migration and capture theory is not able to reproduce all the observed features of Plutinos. The most difficult challenge is their high inclinations $(i)$ up to about $40^{\circ}$ (Fig. \ref{observed}), or probably even larger values. Several extremely inclined KBOs have been reported, such as the scattered disk objects\footnote{The scattered disk objects comprise KBOs on highly eccentric orbits (up to $e=0.9$) with perihelia beyond 30\,AU, and they also have moderate to large inclinations.} Eris with $i\approx44^{\circ}$ (Brown et al.  2005) and 2004 XR190 with $i\approx47^{\circ}$ (Allen et al. 2006; Gomes 2011); the centaurs\footnote{The centaurs are cometary objects who have perihelia and semimajor axes between the orbits of Jupiter and Neptune. The planet-crossing orbits of centaurs are dynamically unstable, and these bodies are believed to be a transient population between the scattered disk objects and the Jupiter-family comets.} 2002 XU93 with $i\approx78^{\circ}$ (Elliot et al. 2005) and 2008 KV42 moving on a retrograde orbit with $i\approx104^{\circ}$ (Gladman et al. 2009). From the La Silla-QUEST KBO Survey, an unusual member of this high-inclination population, 2010 WG9 with perihelion $q\approx18.8$\,AU, $a\approx53.8$\,AU, and $i\approx70^{\circ}$ has been discovered (Schwamb et al. 2011). As the high-$i$ objects spend only a tiny fraction of their time near the ecliptic plane, surveys for KBOs have poor sensitivity to them. With this strong observational bias, we cannot rule out a much wider inclination distribution of Plutinos compared to that of observed ones. Several studies have been done to explore the inclination dependence of the orbital dynamics of Plutinos, and the long-term stability is confirmed for $i\le35^{\circ}$ (Duncan et al. 1995; Morbidelli 1997; Melita \& Brunini 2000; Nesvorn\'{y} \& Roig 2000; Lykawka \& Mukai 2007; Tiscareno \& Malhotra 2009). The purpose of this paper is to extend previous works to a somewhat wider range of inclinations, up to $i=40^{\circ}$ according to currently observed Plutinos. Furthermore, potential Plutinos with $40^{\circ} < i \le 90^{\circ}$ have also been taken into account, in order to obtain a complete view of Plutinos on all prograde orbits. As we shall see, Plutinos may be classified into two groups in terms of a critical inclination about $48^{\circ}$ based on their dynamical behaviors.

In the first resonance sweeping models (Malhotra 1993, 1995; Gomes 1997), the primordial planetesimal disk was naturally assumed to be in dynamically cold orbital conditions (low values of inclination and eccentricity), because it represented the early Solar system after the formation of the planets. In this way, it was possible to explain the origin of the eccentricities of Pluto and the other few Plutinos locked in the 2:3 NMMR known at that time. And for Pluto's large inclination ($\sim17^{\circ}$), Malhotra (1998) proposed a hypothetical excitation process: an inclination increase by the sweeping $\nu_{18}$ secular resonance (the frequency of a planetesimal's nodal precession coincides with that of Neptune), followed by a trapping into the 2:3 NMMR, and later an extra inclination excitation due to the Kozai mechanism (the libration of the argument of perihelion of a planetesimal). Gomes (2000) examined the similar scenario in more detail, showing that most of the inclination increase is produced by the effective $\nu_{18}$ secular resonance before the 2:3 NMMR arrives. However, in such a scenario the maximum $i$ cannot achieve higher than $20^{\circ}$, while the eccentricity $e$ undergoes large increase due to the $\nu_{8}$ secular resonance (the frequency of a planetesimal's perihelion precession matches that of Neptune), inconsistent with the current $e$-$i$ distribution of observed Plutinos (Fig. \ref{observed}). Later, Gomes (2003) suggested that the high-inclination Plutinos are originated from eccentric Neptune-crossing orbits around 25\,AU by dynamical scatterings, and have their orbits circularized on the path to the 2:3 NMMR; while the population with low $i$ is expected to have been formed in a more distant area from the Sun. Thus, the various formation environments such as the temperature and the pressure could result in different surface colors of the high- and low-inclination groupings (Trujillo \& Brown 2002). But in a recent paper, Sheppard (2012) reported that Plutinos have no apparent inclination-color correlation from observations.

In the framework of the $\mathit{Nice}$ model, a planetesimal disk with its outer edge interior to the position of the 1:2 NMMR at the time of the planetary instability is considered as the source of the high-inclination KBOs (Tsiganis et al. 2005; Levison et al. 2008). There was a time when Uranus and Neptune experienced a short but violent phase, the inclinations of disk particles would be stirred up dramatically. Meanwhile, if the eccentricity of Neptune exceeded 0.15, the NMMRs interior to the 1:2 NMMR overlapped one another. Thus the region from Neptune's orbit to the 1:2 NMMR became a chaotic sea, throughout which the high-inclination population could swim outwards freely, and some of them were captured by the 2:3 NMMR in the following evolution. The $e$-$i$ distribution of survivors in the 2:3 NMMR resembles that of observed Plutinos very well, regardless of a deficit of objects with $i>30^{\circ}$.

Some other mechanisms have also been proposed to account for the inclination excitation of KBOs.  Morbidelli \& Valsecchi (1997), Petit, Morbidelli \& Valsecchi (1999) and Lykawka \& Mukai (2008) showed that large planetesimals were likely scattered by one of the Jovian planets, then they could have pumped up $e$ and $i$ of the vast majority of KBOs. Ida, Larwood \& Burkert (2000) and Kobayashi, Ida \& Tanaka (2005) suggested that if an early encounter with a stellar of one solar mass and a perihelion distance on the order of 100\,AU occurred in the past, most objects in the primordial Kuiper belt could acquire high values of $i$, which are limited by the inclination of the passing star relative to the midplane of planetesimal disk.  Chiang et al. (2007) argued that during the era of planetary system formation, a few Neptune-mass oligarchs existed between 20\,AU and 40\,AU may have stimulated the KBOs' orbits before they were finally removed. Following Nagasawa \& Ida (2000), Li, Zhou \& Sun (2008) pointed out that, the KBOs can be excited to $i\sim30^{\circ}$ via the sweeping $v_{15}$ secular resonance (the frequency of a planetesimal's perihelion precession is close to that of Jupiter).

Taken in total, many works focus on how to raise $i$ under possible perturbations, but there is also evidence indicating that the planetesimal disk has been dynamically heated prior to the onset of the planet migration. We know that, for objects locked in the 2:3 NMMR with $i\gtrsim10^{\circ}-15^{\circ}$, their $i$ cannot be modified significantly during both the migration epoch  (Gomes 2000, 2003; Li et al. 2011) and the later Gyr time evolution near 39.4\,AU (Volk \& Malhotra 2011). This suggests that Plutinos should have already possessed their high $i$ before they were hijacked into the 2:3 NMMR, and approximately conserved $i$ values from that point forward. In addition, Wiegert et al. (2003) examined the capture of particles into the 2:3 NMMR by the effect of Neptune's in situ accretion, and they showed that Pluto should have arisen from an initial high-inclination ($\sim25^{\circ}$) orbit. In this paper, we present new insights on Neptune's migration into a three-dimensional planetesimal disk, and attempt to get some deterministic results about the role of $i$ in the origin of high-inclination Plutinos. Since it is natural to restrict the maximum initial $i$ of primordial planetesimals to the possible $i$-range of stable Plutinos, the study of a previously inclined population may provide novel constraints on the early history of the outer Solar system.

The present manuscript aimed at the dynamics of the high-inclination Plutinos is planed as follows. In Section 2, we describe some particular features of the 2:3 NMMR for inclined orbits. And we also derive the minimum of the resonant amplitude as a function of $i$. Section 3 discusses the stability of high-inclination objects in the 2:3 NMMR for times up to the age of the Solar system, and shows that the stable region covers the whole inclination space of $i\le90^{\circ}$. Based on this argument, in Section 4 we investigate in detail the evolution of test particles with initial $i$ as high as $90^{\circ}$ in the planet migration and resonance sweeping model. Finally, the conclusions and discussion are given in Section 5.


\section[]{The resonant features}

Considering the restricted three-body model that consists of the Sun, Neptune and a Plutino, we can write the disturbing function in the heliocentric coordinates (Murray \& Dermott 1999)
\begin{equation}
\mathcal{R}=\frac{Gm_N}{|\mathbf{r}_N-\mathbf{r}|}-Gm_N\frac{\mathbf{r}\cdot\mathbf{r}_N}{r_N^3},
\label{df}
\end{equation}
where $G$, $m$ and $\mathbf{r}$ are the constant of gravitation, the mass and  the position vector, respectively. The subscript $N$ refers to Neptune and no subscript indicates Plutino. We will assume in this section that Neptune is on a circular and zero-inclination orbit around the Sun. 

We regard a small body as a Plutino if the critical argument of the 2:3 NMMR
\begin{equation}
\sigma=2\lambda_N-3\lambda+\varpi
\label{angle}
\end{equation}
librates, where $\lambda$ and $\varpi$ are the mean longitude and the longitude of perihelion. Since the change rate of $\sigma$ is slow, we can average the disturbing function (\ref{df}) over the fast variable $\lambda_N$ to remove it. For the Plutino at the location of nominal resonance with Neptune, i.e., $a=(3/2)^{2/3}a_N$, we have 
\begin{equation}
R(\sigma)=\frac{1}{6\pi}\int_{0}^{6\pi}\mathcal{R}\left(\lambda_N, \lambda(\lambda_N,\sigma)\right)d\lambda_N
\label{df2}
\end{equation}
for fixed values of $e, i, \varpi, \sigma$ and the longitude of ascending node $\Omega$, where $\lambda=\lambda(\lambda_N,\sigma)$ is deduced from equation (\ref{angle}). The integration interval of equation (\ref{df2}) is adopted to be the time spent between successive conjunctions of the Plutino and Neptune, i.e.,  3 times the Neptune's orbital period of $2\pi$.

In order to evaluate $R(\sigma)$ for the 2:3 NMMR, we follow the technique developed by Gallardo (2006a, 2006b).  First of all, since the resonant angle $\sigma$ is invariant to changes in Aries in the circular restricted three-body problem, Aries is taken as the ascending node. Then $\Omega$ could be set to be 0, and $\varpi$ equals to the argument of perihelion $\omega$. Secondly, the slowly evolving $\sigma$ is assumed to be constant during the above integration interval, which is much shorter than the resonant period, then $\mathcal{R}(\lambda_N, \lambda)$ could be calculated from equation (\ref{df}). Finally, we obtain the averaged disturbing function $R(\sigma)$, which depends only on a given set of fixed values of  ($e, i, \omega, \sigma$).

\begin{figure}
  \hspace{-0.7cm}
  \centering
  \includegraphics[width=9cm]{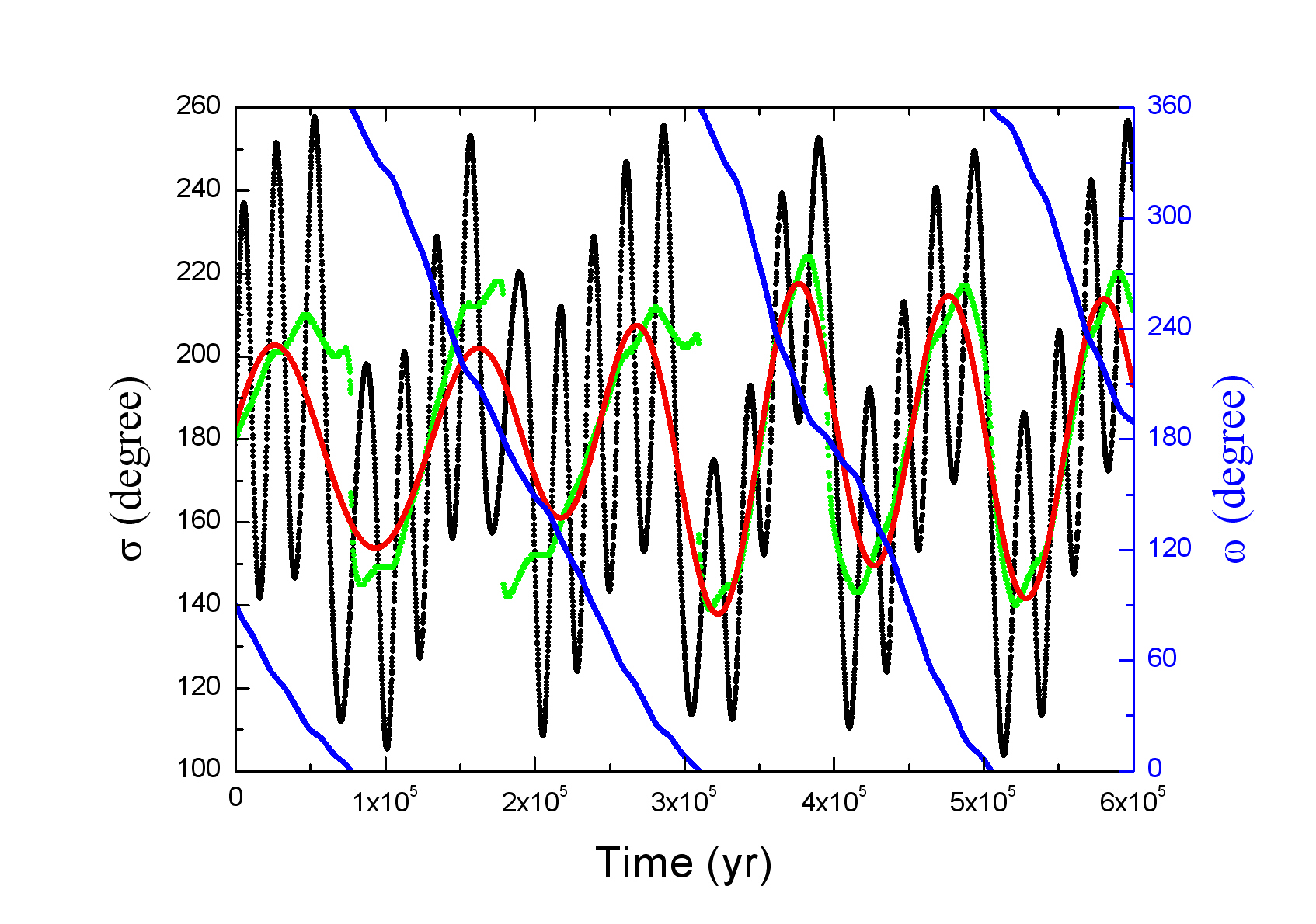}
  \caption{Time evolution of the resonant angle $\sigma=2\lambda_N-3\lambda+\varpi$ (black curve) and the argument of perihelion $\omega$ (blue curve) for a typical inclined Plutino with $i=20^{\circ}$. The red curve represents the secular behavior of $\sigma$, and the green curve corresponds to the temporal variation of the SLC (see text for the definition).}
  \label{e005i30}
\end{figure}

Next, we will determine the libration center of the 2:3 NMMR using $R(\sigma)$ for a series of $\sigma$ between 0 and $2\pi$. Here we refer to such libration center as the special libration center (SLC). To do this we introduce Lagrange's planetary equations, which give the time derivative of the semimajor axis
\begin{equation}
\frac{da}{dt}=\frac{2}{na}\cdot\frac{\partial R}{\partial \lambda}=\frac{2}{na}\cdot\frac{\partial R}{\partial \sigma}\cdot\frac{\partial \sigma}{\partial \lambda}\propto\frac{\partial R}{\partial \sigma},
\end{equation}
where $n$ is the mean motion.  As $da/dt$ is proportional to ${\partial R}/{\partial \sigma}$, the minimum of $R(\sigma)$ defines the SLC for a specific set of ($e, i, \omega$). Gallardo (2006b) described the general behavior for the 2:3 exterior resonance:  there are two SLCs at $\sigma=0^{\circ}$ and $180^{\circ}$ for low-inclination orbits; but for high-inclination orbits, the position of the SLC strongly depends on $\omega$. 

It is important to note that the change of the SLC has a direct influence towards the resonant amplitude. Here we designate the mean value of $\sigma$ during the evolution as the general libration center (GLC), and the resonant amplitude as the maximum deviation from the GLC. By definition, a particle inhabits the exact resonance if its resonant amplitude is equal to zero. However, this resonant configuration can never be achieved so long as the particle moves on an inclined orbit. Using the direct numerical integration, we illustrate in Fig. \ref{e005i30} the evolution of a typical inclined Plutino P1 started at the SLC $\sigma=180^{\circ}$ for initial $i=20^{\circ}$, $e=0.05$ and $\omega=90^{\circ}$. In a period of $6\times10^5$ yr, the resonant amplitude of P1 is as large as $80^{\circ}$, with the GLC at $180^{\circ}$.   

\begin{figure}
  \centering
  \includegraphics[width=9cm]{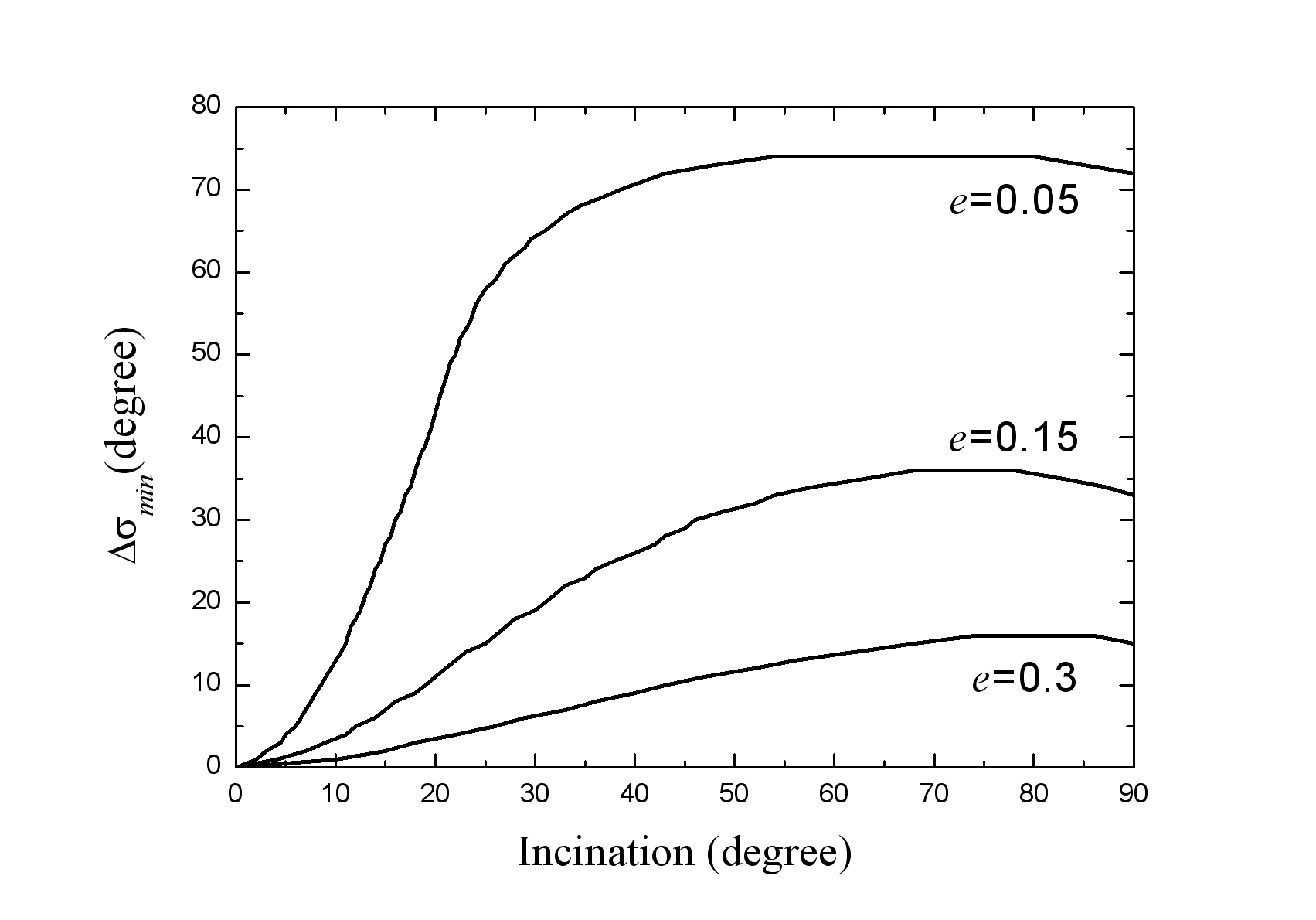}
  \caption{The resonant amplitude's lower limit ($\Delta\sigma_{min}$) for orbits with inclinations from $0^{\circ}$ to $90^{\circ}$, at three representative eccentricities of $e=0.05$, 0.15 and 0.3. Note that the high-inclination Plutinos always have non-zero resonant amplitudes, thus they will never settle at the exact 2:3 NMMR.}
  \label{amplitude}
\end{figure}

As shown in Fig. \ref{e005i30}, the temporal variation of $\sigma$ of P1 is a composition of the secular and short-period behaviors. So we performed an FFT low pass filter to $\sigma$ to extract the low-frequency term $\bar{\sigma}$, as indicated by the red curve in Fig. \ref{e005i30}. We find that the libration period of $\bar{\sigma}$ coincides very well with half the circulation period of $\omega$. This can be understood by looking at the time derivative of the resonant angle 
\begin{equation}
\dot{\sigma}=2\dot{\lambda}_N-3(\dot{\omega}+\dot{M})+\dot{\omega}=2n_N-3n-2\dot{\omega},
\label{dangle}
\end{equation}
which has two fundamental frequencies $f_1=2n_N-3n$ and $f_2=-2\dot{\omega}$. Because $\left\vert f_2\right\vert$ is much smaller than $\left\vert f_1\right\vert$ for inclined Plutinos, the variation of $\omega$ is responsible for the secular behavior of $\sigma$, i.e., the temporal evolution of $\bar{\sigma}$. Keeping in mind that the SLC is also sensitively dependent on $\omega$, we calculate it for specific values of $e, i$ and $\omega$ from P1's instantaneous orbit. The temporal behavior of the SLC is plotted as the green curve in Fig. \ref{e005i30}. The high-degree overlap between red and green curves demonstrates that the secular behavior $\bar{\sigma}$ could be essentially identical to the variation of the SLC, which is not a constant anymore for high-inclination Plutinos. The amplitude of the SLC is about $41^{\circ}$ corresponding to a low frequency $f_2$, and the oscillating of $\sigma$ superposed on the SLC with a high frequency $f_1$ has an amplitude about $39^{\circ}$. These two parts comprise the global deviation of the resonant angle $\sigma$ from the GLC at $180^{\circ}$, namely what is usually called ``resonant amplitude''.

In view of the above, for high-inclination Plutinos the amplitude of the SLC predicts the resonant amplitude's lower limit, denoted by $\Delta\sigma_{min}$. Given three representative eccentricities of $e=0.05$, 0.15 and 0.3 (covering the observational range), we estimated the values of  $\Delta\sigma_{min}$ for Plutinos with $i\le90^{\circ}$ (see Fig.  \ref{amplitude}). As expected, $\Delta\sigma_{min}$ is zero when $i=0^{\circ}$ and shifts to larger value with increasing $i$, reaching a maximum of $\sim75^{\circ}$ under $i\sim80^{\circ}-90^{\circ}$ for $e=0.05$. Along with the increase of $e$, the variation of the SLC becomes less and less pronounced. When $e$ exceeds 0.3, the values of $\Delta\sigma_{min}$ are always on the order of smaller than $20^{\circ}$ for all prograde Plutinos. It is necessary to stress that, the exact 2:3 NMMR with zero-amplitude libration will never arise unless in the two-dimensional planar three-body model, where the SLC is fixed at $0^{\circ}$ or $180^{\circ}$.  In addition, we would like to mention that low-inclination Plutinos librating around $\sigma=0^{\circ}$ will lose stability if the perturbations of any other Jovian planet besides Neptune are taken into account.   

In the following of this paper, we will focus on high-inclination Plutinos with $i\ge10^{\circ}$. For consistency with previous literatures, we hereafter denote the libration center of the 2:3 NMMR by $\sigma_0=180^{\circ}$ (i.e., the GLC), and the resonant amplitude by $A_{\sigma}=\sigma_{max}-\sigma_0$.


\section{The long-term stability}

To update the possible upper-$i$ cutoff of Plutinos, we must first analyze the dynamical stability of high-inclination candidates populating in the 2:3 NMMR. This section presents our results of numerical integrations of Plutino-type orbits (i.e., in the 2:3 NMMR with $a\sim39.4$\,AU and the resonant angle ${\sigma}$ librating) with $i$ as large as $90^{\circ}$ for the age of the Solar system ($4\times10^9$ yr). As we know by far, there is just one similar work considering such extremely inclined orbits (up to $i=90^{\circ}$) done by Duncan et al. (1995); however, the duration of their integrations is relatively short ($10^9$ yr), and the initial eccentricities are restricted in the range $e\le 0.1$, where inclined orbits are more easily destabilized by secular resonances (Nesvorn\'{y} \& Roig 2000).

\begin{figure}
  \centering
  \includegraphics[width=9cm]{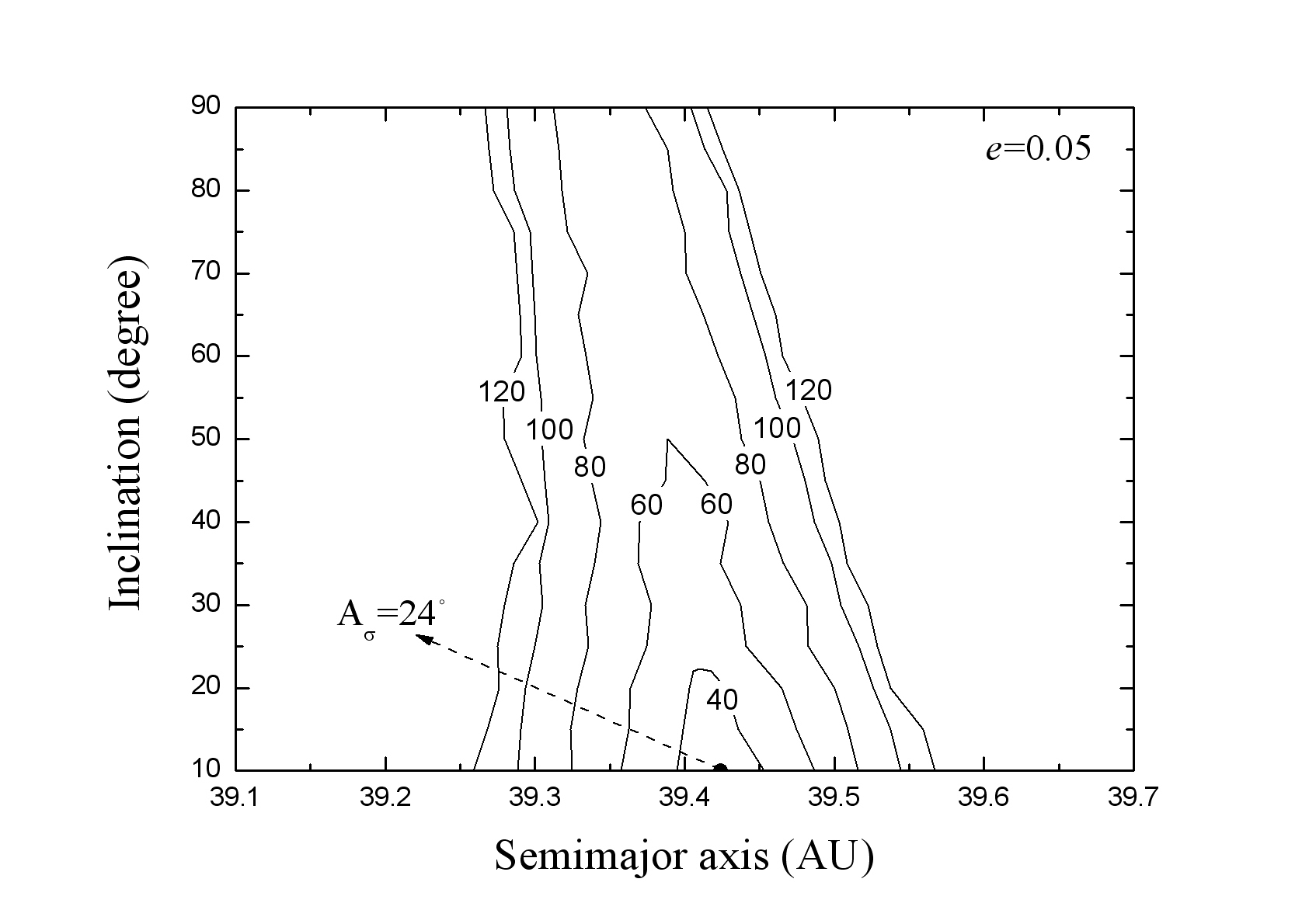}
  \caption{The contours of the resonant amplitude $A_{\sigma}$ of the 2:3 NMMR on the initial $(a, i)$ plane for the eccentricity of $e=0.05$. The minimum of $A_{\sigma}$ is $24^{\circ}$ at $i=10^{\circ}$.}
  \label{amplitude2}
\end{figure}

In our numerical model, the Solar system consists of the Sun with the masses of the terrestrial planets added, four Jovian planets and a number of test particle Plutinos. The planets' masses, initial positions and velocities are taken from DE405 (Standish 1998). Initially, test particles are uniformly spaced between 39\,AU and 40\,AU with $i$ ranging from $10^{\circ}$ to $90^{\circ}$ in steps of $10^{\circ}$, still for $e=0.05$, 0.15 and 0.3. The initial angles are chosen such that all particles start from $\sigma=\sigma_0=180^{\circ}$, $\omega=90^{\circ}$ (the libration center of $\omega$ associated with the Kozai mechanism) and $\Omega=\Omega_P$ (the longitude of ascending node of Pluto). The massless particles are gravitationally affected by, but not affecting the other bodies. In this paper we employ the swift\_rmvs3 symplectic integrator developed by Levison \& Duncan (1994). We use a time step of 0.5 year, which is about 1/24 of the shortest orbital period (Jupiter) in our model.

According to Morbidelli (1997), the dynamical structure of the 2:3 NMMR at small inclination can be simply characterized by the resonant amplitude $A_{\sigma}$, which determines the regular, chaotic and transition regions. This stability criterion warrants an extension to highly inclined orbits. Thus, we have numerically calculated and recorded $A_{\sigma}$ as a function of initial $a$ and $i$, by performing a series of high-resolution runs with $\Delta a=0.001$\,AU and $\Delta i=5^{\circ}$ for $10^6$ years. A typical example of the refined contours of $A_{\sigma}$ has been shown in Fig. \ref{amplitude2} for $e=0.05$. From this case, we again prove that the inclined Plutinos would never stick to $\sigma_0$ with $A_{\sigma}=0^{\circ}$. The minimum of $A_{\sigma}$ at a specific $i$ is always a bit larger than the corresponding lower limit $\Delta\sigma_{min}$ obtained by our analysis in Section 2. With different initial $\omega$ or $\Omega$, we performed additional calculations for the case of $e=0.05$. We find that, as long as each of these two angles is the same for all orbits starting from $\sigma=180^{\circ}$, the shape of $A_{\sigma}$ contours is equivalent to Fig. \ref{amplitude2}. We also recall from Morbidelli et al. (1995) that the 2:3 NMMR has a very complex dynamics since some secular resonances are located inside, as we discuss later.

\begin{figure}
  \centering
  \begin{minipage}[c]{0.5\textwidth}
  \centering
  \hspace{-0.8cm}
  \includegraphics[width=9cm]{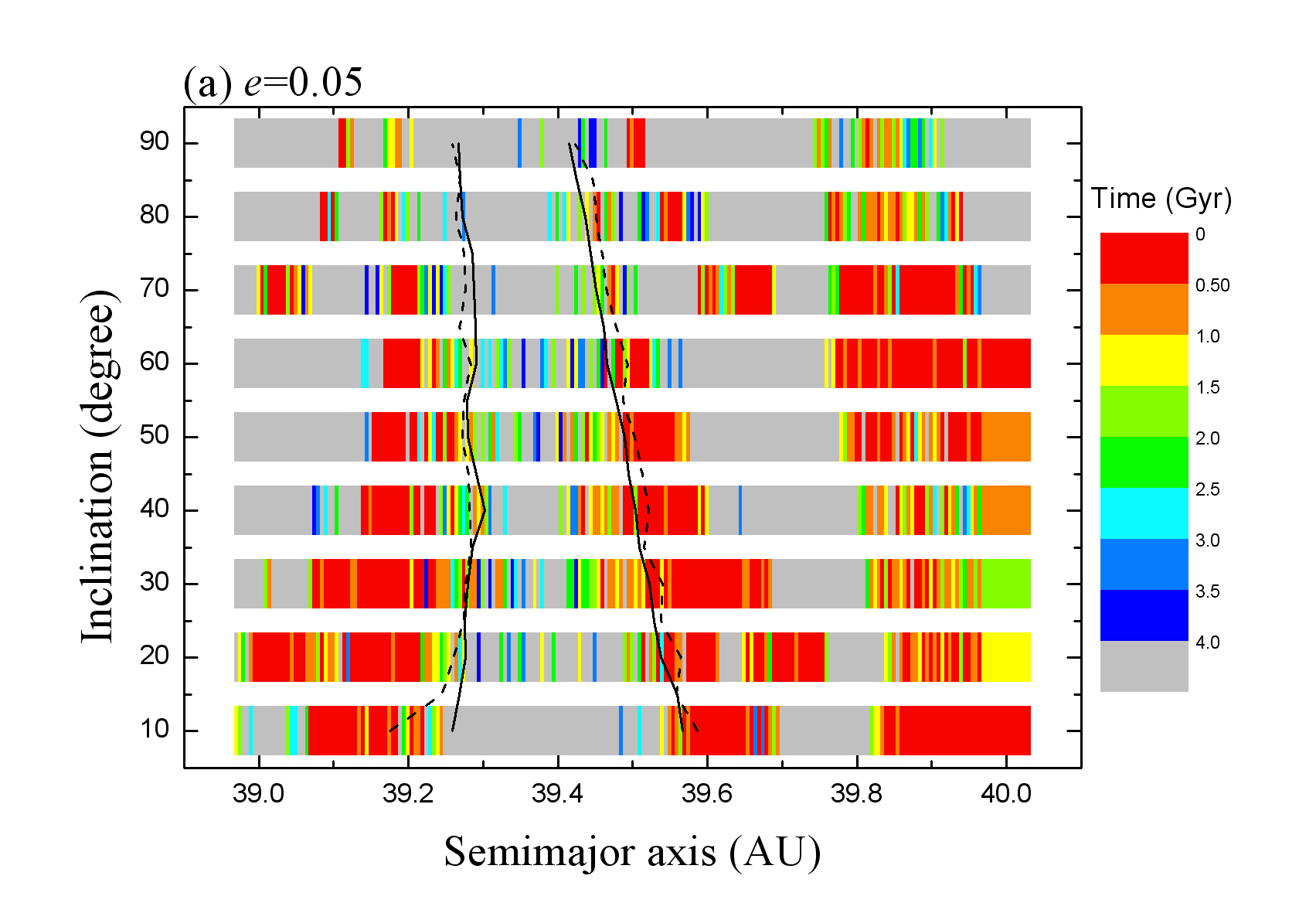}
  \end{minipage}
  \begin{minipage}[c]{0.5\textwidth}
  \centering
  \hspace{-1.5cm}
  \includegraphics[width=9cm]{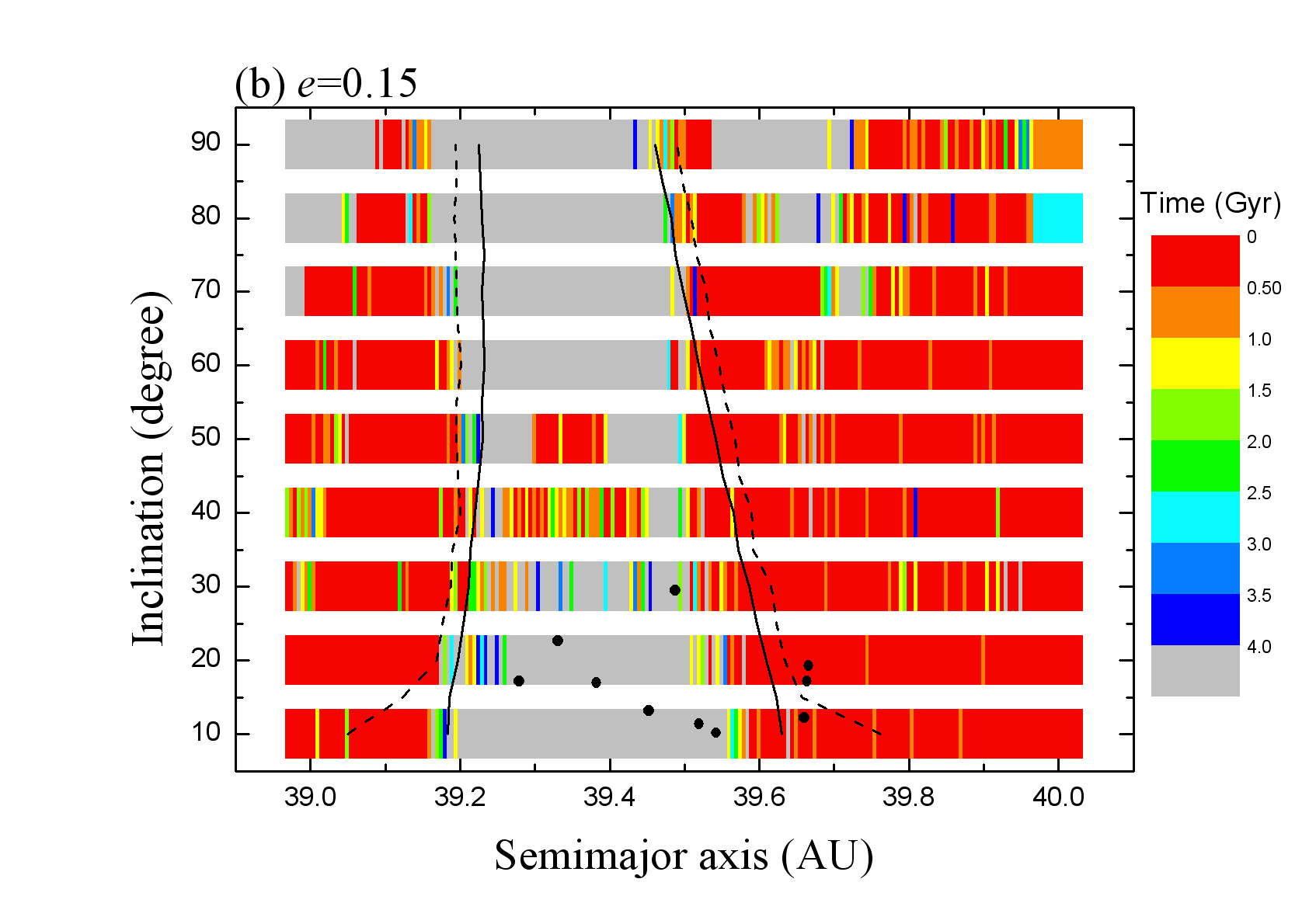}
  \end{minipage}
  \begin{minipage}[c]{0.5\textwidth}
  \centering
  \hspace{-1.5cm}
  \includegraphics[width=9cm]{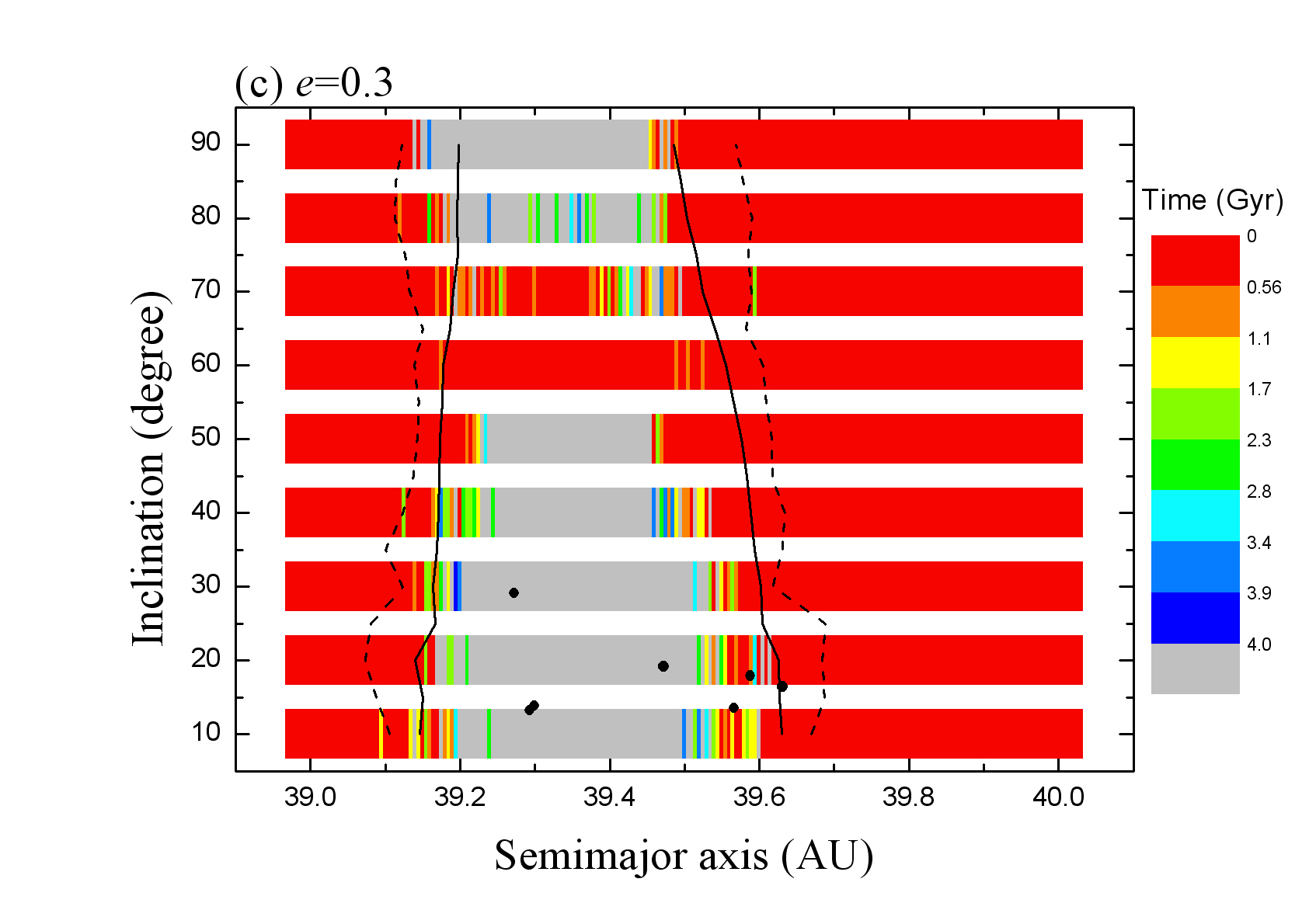}
  \end{minipage}
  \caption{The dynamical lifetime on the initial $(a, i)$ plane for test particles spaced between 39\,AU and 40\,AU in the $4\times10^9$ yr numerical simulations: (a) $e=0.05$, (b)  $e=0.15$, (c) $e=0.3$. The grey strips represent particles that can survive for the length of our integration. The dashed lines denote the separatrices of the 2:3 NMMR (i.e., $A_{\sigma}=180^{\circ}$). The contour curves of $A_{\sigma}=120^{\circ}$ (full lines) roughly surround the stable orbits in the 2:3 NMMR, which can extend to inclinations as high as $90^{\circ}$. The dots in panels (b) and (c) indicate the observed Plutinos with $e=0.15\pm0.015$  and $e=0.3\pm0.015$, respectively.}
 \label{stability}
\end{figure}

In the following $4\times10^9$ yr integrations, we reduce the spatial resolution to $\Delta a=0.005$ \,AU and $\Delta i=10^{\circ}$ (i.e., 201 test particle lie between 39\,AU and 40\,AU at each $i$) in order to save the amount of computer time. Even so, this resolution is much higher than that in Duncan et al. (1995) ($\Delta a=0.1$ \,AU), and it is effectual to measure the upper-$i$ cutoff, below which objects could remain stable in the 2:3 NMMR. The dynamical lifetime of each test particle is calculated from our numerical simulations, and it is indicated by the colored strip on the initial $(a, i)$ plane in Fig. \ref{stability}. In the gray region, Plutinos can survive right up to the present day. 

The key result depicted in Fig. \ref{stability} is that the stable motion in the 2:3 NMMR covers the whole inclination-range of $10^{\circ}\le i \le90^{\circ}$. Figure \ref{stability}a could be combined with the middle panel of Fig. 2 from Duncan et al. (1995), i.e., for the case of initial $e=0.05$,  the existence of stable regions near 39.4\,AU at large $i$ up to $90^{\circ}$ has been further confirmed in our long-time integrations over 4 Gyr. The main discrepancy is that the 2:3 NMMR is fully unstable at $i=30^{\circ}$ for $e=0.05$ in Duncan et al.'s calculations, and it probably results from their lower resolution of $a$. As can be seen by careful examination of Fig. \ref{stability}a, at $i=30^{\circ}$, the widest stable range of $a$ in the 2:3 NMMR is only about 0.06\,AU. While for the coverage in Duncan et al. (1995), the initial spacing between test particles was chosen to be 0.1\,AU, thereby they might leave out the narrow stable bands there. In reality, Tiscareno \& Malhotra (2009) found that stable resonant orbits for Plutinos are visible for the entire range of $5^{\circ}\le i \le 35^{\circ}$ at low $e$. However, would Plutinos with $i$ extending even to $90^{\circ}$ be expected to be discovered? The transport capacity of the sweeping 2:3 NMMR for the high-inclination population will provide further information on this question, as we will consider in the next section.

In Fig. \ref{stability} we plotted the separatrices of the 2:3 NMMR ($A_{\sigma}=180^{\circ}$, dashed lines) and the contour curves of $A_{\sigma}=120^{\circ}$ (full lines). It is suggested by Levison \& Stern (1995) that for Plutinos the orbits with  $A_{\sigma}>120^{\circ}$ are typically unstable over $4\times10^9$ yr. Overall, we can see that the overwhelming majority of stable resonant members generated in our simulations lie well inside the region of $A_{\sigma}<120^{\circ}$. Thus this criterion based on $A_{\sigma}$ can roughly be used to judge the long-term stability of all orbits with $i \le 90^{\circ}$ in the 2:3 NMMR. It is important to notice that the critical value $A_{\sigma}\sim120^{\circ}$ is larger than the prescribed lower limit of $A_{\sigma}$ for any $i$, so the existence of stable motion is permitted. Moreover, on closer inspection of Fig. \ref{stability}, there are still some complex dynamical structures in the neighborhood of the resonance center at $\sim39.4$\,AU (i.e., the location of nominal 2:3 NMMR), which is not generally simply covered by the simulated long-lived Plutinos as in the case of $i=10^{\circ}$.

For the case of small $e$	($\sim0.05$), the central stable bands of the 2:3 NMMR extend to $i=90^{\circ}$ (Fig. \ref{stability}a). Out of these narrow stable regions near the resonance center, the stable and unstable regions interlace with one another. This is principally due to the appearance of the Kozai mechanism. Besides the effect of perihelion protection against Neptune, the Kozai mechanism may also bring strong orbital instability via transferring $i$ to $e$. We carefully examined the unstable orbits inside the 2:3 NMMR for $e=0.05$, and confirmed that most of them underwent the temporary libration of $\omega$. Such kind of instability might emerge for $e$ under a certain limit $\sim0.1$. Actually, as shown in Fig. \ref{observed}, none of observed Plutinos experiences the Kozai mechanism in the region $e<0.1$.

Apart from the instability associated with the Kozai mechanism for the case of small $e$, we discover that the $\nu_8$ secular resonance could also contribute to the increase in $e$ for Plutinos with $i$ as large as $70^{\circ}$, leading to close encounters with planets. However, the $\nu_8$ secular resonance turns out to be a rare occurrence, and only provides a few escaping routes in our simulations. Frequently, the $\nu_8$ secular resonance is responsible for the slow-diffusing orbits by acting together with the Kozai mechanism. While the $\nu_{18}$ secular resonance is limited to $i<10^{\circ}$ (Morbidelli et al. 1995), it has no influence on the stability of our highly inclined samples in the 2:3 NMMR.

\begin{figure}
  \hspace{-8mm}
  \centering
  \includegraphics[width=9cm]{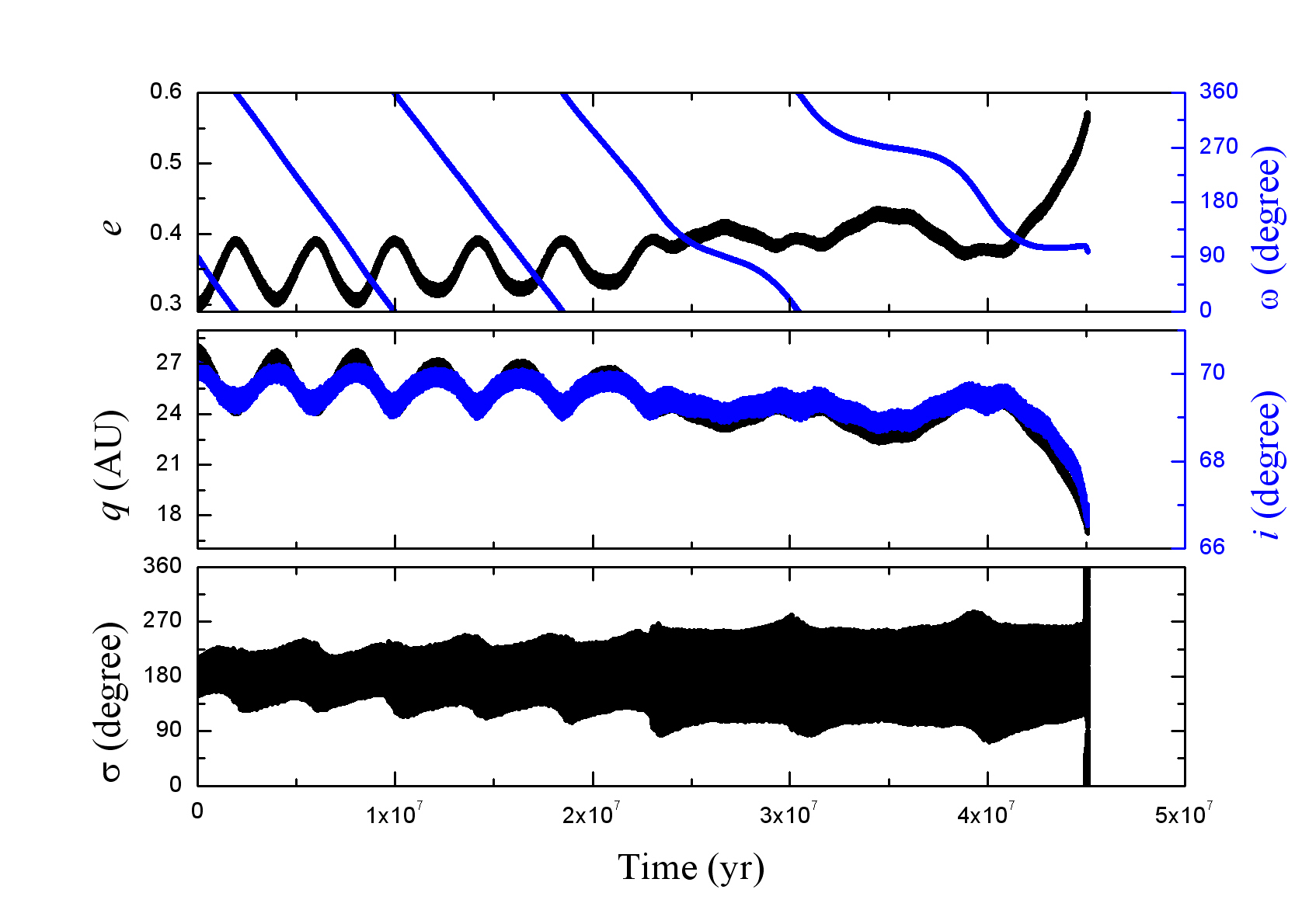}
  \caption{ Example of the escape of a high-inclination test particle from the central region of the 2:3 NMMR due to the Kozai mechanism. This particle with initial  $a=39.4$\,AU came from the case of $e=0.3$ at $i=70^{\circ}$ (see Fig. \ref{stability}c). At about $4.1\times10^7$ yr, $e$ starts to increase significantly as the slow variation of $\omega$ around $90^{\circ}$ is evinced (upper panel). The corresponding decreases in perihelion distance $q$ and $i$ are plotted in the middle panel. Around $4.5\times10^7$ yr, the 2:3 NMMR is broken (lower panel), and then the particle is scattered away by Neptune. }  
\label{encounter}
\end{figure}

For the case of moderate to large $e$ ($\sim0.15-0.3$), the stable regions inside the 2:3 NMMR exhibit different coverage with the change of $i$ (Fig. \ref{stability}b, c): 

(1) When $i$ is below a certain limit $i_1$, a wide neighborhood of the resonance center is almost covered by the long-lived orbits. Thus the principal part of the stable region can be simply scaled by a specific value of $A_{\sigma}$. Based on our detailed investigation, we propose that the certain inclination $i_1$ shifts to larger value with increasing $e$. For Plutinos with $i<i_1$ in our simulations, once trapped into the Kozai mechanism, they would be under the long-lasting $\omega$-libration phase and remain stable until the end of the integration.

(2) For orbits with $i_1 < i < i_2$ ($i_2$ is another certain limit), the most distinguishable feature is the instability across the resonance center. We examined the evolution of Plutinos in this $i$-range and found that the status of the Kozai mechanism is mainly temporary, resulting in the destruction of the high-inclination population rather than the protection, similar to the case of small $e$. Here, the upper limit $i_2$ also increases with larger $e$.

Fig. \ref{stability}c shows a typical dynamical structure at $i=70^{\circ}$. Most test particles in the central region of the 2:3 NMMR evolve into the Kozai mechanism with slow variation of $\omega$ in less than 1 Gyr, and have their $e$ pumped up to Uranus-crossing values. Then they would suffer strong perturbations from close encounters with Neptune or Uranus, and be cleared away from the Kuiper belt.  A example started at $a=39.4$\,AU of this process is given in Fig. \ref{encounter}. Besides, in Fig. \ref{stability}c we note that there are a handful of objects that could remain in the 2:3 NMMR on timescales of $2-3$ Gyr, while the similar $e$-excitation and elimination event will also occur. Although the stability near the resonance center is gone, some Plutinos located at $ A_{\sigma}>95^{\circ}$ could still survive after 4 Gyr evolution, residing in several narrow (gray) regions.  

It is worth mentioning that there exist intermediate zones near $i_1$ and $i_2$, where the Kozai mechanism has two faces.  For instance, at $i=30^{\circ}$ ($\sim i_1$) and $e=0.15$ (Fig. \ref{stability}b), a part of $\omega$-librators comprise the Plutinos that are stable for the length of our integration; while some objects evade the libration of $\omega$ and then become unstable. And the similar examples can also be found for orbits with $i=80^{\circ}$ ($\sim i_2$) and $e=0.3$.

An unstable gap throughout the entire 2:3 resonant region at $i=60^{\circ}$ for $e=0.3$ can be seen in Fig. \ref{stability}c, but it is not a concern in this paper. Our interest is in offering the possibility of the existence of high-inclination Plutinos, and the stable ones with $i=60^{\circ}$ have been confirmed at other $e$ values (0.05 and 0.15). The same consideration also applies for other similar unstable regions at $i=70^{\circ}$ for $e=0.3$ and $i=40^{\circ}$ and $50^{\circ}$ for $e=0.15$.

(3) At $i_2<i\le90^{\circ}$, Plutinos can hardly experience the (temporary) Kozai mechanism. Once again, the center of the 2:3 NMMR is surrounded by stable orbits, which are constrained only by $A_{\sigma}$.

Here we refer $i_1$ and $i_2$ to initial inclinations of test particles, thus it is straightforward to qualitatively portray the general trend toward stability for high-inclination Plutinos. Based on the former results, given $\Delta i=10^{\circ}$, their tentative values are estimated to be $i_1\sim30^{\circ}$ and $i_2\sim50^{\circ}$ for $e=0.15$; $i_1\sim60^{\circ}$ and $i_2\sim80^{\circ}$ for $e=0.3$. These two limits are only certain for a specific initial $e$, and a great deal of computing is needed to determine the more precise $i_1(e)$ and $i_2(e)$ using a much higher $i$-resolution in the future. 

For a quick reference, in panels (b) and (c) of Fig. \ref{stability}, we superimposed the positions of the observed Plutinos (dots) with $e=0.15\pm0.015$ and $e=0.3\pm0.015$,  respectively. These objects are selected according to their $e$-values from Fig. \ref{observed} with $i\ge10^{\circ}$. We can see that most of them are embedded in or very close to the long-term stable (grey) regions. However, there are a few observed Plutinos lying in the very unstable (red) area, but actually they are moving on stable orbits. This inconsistency could stem from the fact that their instantaneous angular variables cannot fully satisfy the initial conditions $\sigma=180^{\circ}$, $\omega=90^{\circ}$ and $\Omega=\Omega_P$ of test particles used in our stability calculations. It should be remarked that, there is no Plutinos plotted in Fig. \ref{stability}a due to the lack of the observation data for orbits with $e\le0.066$ and $i\ge10^{\circ}$ (see Fig. \ref{observed}).


\section{Resonant capture and orbital evolution}

As shown above, the inclinations of potential Plutinos could be as high as $90^{\circ}$. Then it is of great significance to investigate whether primordially inclined planetesimals originated from a smaller heliocentric distance could be captured into the 2:3 NMMR and transported outwards to $a\approx39.4$\,AU during Neptune's migration. Up to now, a number of works have been done to study the probability of resonant capture and retention as a function of the initial $a$ and $e$ of test particles, or the migration speed of Neptune, and so on (Melita \& Brunini 2000; Friedland 2001; Chiang \& Jordan 2002). In this section we would explore the unknown role of particle's initial $i$.

\subsection{The migration model}

\begin{table*}
\centering
\begin{minipage}{130mm}
\caption{The statistics of captured Plutinos for initial inclinations from $10^{\circ}$ to $90^{\circ}$ within the planet migration model (see Section 4.2 for detailed descriptions of the symbols).}      
\label{tab:plutinos}
\begin{tabular}{c c c c c c c l  | c}        
\hline                 
$i_0(^{\circ})$       &         $R_{2:3}(\%)$      &     $K_{90}(\%)$    &     $K_{270}(\%)$    &       $e$       &    $i(^{\circ})$     &    Min$A_{\sigma}(^{\circ})$     &     $N_{7:11}$   &\\

\hline

10   &      58.5      &     24.8      &     30.0       &       0.12-0.29      &     3-15       &    13   &    --   &   \multirow{4}{*}{Population \Rmnum1} \\

20   &      53.5      &     29.0      &     25.2       &       0.06-0.27      &    14-24      &    16   &    2 ~(1\%)  &   {} \\
 
30   &      23.5      &     38.3       &    19.1       &       0.06-0.35	   &     25-32     &    29   &    5 ~(2.5\%)   &   {} \\
   
40   &      10.5      &     42.9       &    28.6       &       0.18-0.22	   &     34-40     &    36   &    3 ~(1.5\%) &   {} \\

\hline \hline

50   &        9.5      &       5.3       &     15.8       &       0.16-0.26	   &     45-49     &    48   &    --   &   \multirow{5}{*}{Population \Rmnum2} \\

60   &        16       &         0        &        0	        &       0.19-0.30	   &     57-59     &    41   &    --   &   {} \\

70   &       17.5     &         0        &        0          &       0.21-0.32     &      67-68    &    38   &     --   &   {} \\

80   &         31       &         0        &       0          &       0.20-0.38	   &      76-77    &    30   &    --   &   {} \\

90   &         31       &         0        &       0	        &       0.19-0.35	   &      86-87    &    28   &    --   &   {} \\
\hline
\end{tabular}
\end{minipage}
\end{table*}

The radial migration scenario of Jovian planets is described by a simple time variation of their semimajor axes (Malhotra 1995)
\begin{equation}
  a_p(t)=a_p^{(f)}-\Delta a \exp(-t/\tau),
\label{eq:variation}
\end{equation}
where $a_p^{(f)}$ is each planet's current semimajor axis, $a_p(t)$ is the value at epoch $t$, and $\tau$ is the migration timescale. The amplitudes $\Delta a_p$ of the migration are adopted to be $-0.2$, 0.8, 3.0 and 7.0\,AU for Jupiter, Saturn, Uranus and Neptune, respectively. The smooth migration could be mimicked by adding an artificial force on each planet along the direction of the orbital velocity $\mathbf{\hat{\nu}}$, as
\begin{equation}
  \Delta \mathbf{\ddot{r}} =\frac{\mathbf{\hat{\nu}}}{\tau}
  \left\{\sqrt{\frac{GM_\odot}{a_p^{(i)}}}-\sqrt{\frac{GM_\odot}{a_p^{(f)}}} \right\}
  \exp\left(-\frac{t}{\tau}\right),
\end{equation}
where $a_p^{(i)}=a_p^{(f)}-\Delta a_p$ is the initial semimajor axis of the planet at $t=0$, and $M_\odot$ is the mass of the Sun.

In order to highlight the effects of $i$ on the formation of Plutinos within the planet migration model, we try to diminish the influence of the following factors. Firstly, we consider the dependence of resonant capture on the migration speed. From equation (\ref{eq:variation}) we can estimate the rate of the change of Neptune's semimajor axis $a_N$ as
\begin{equation}
  \dot{a}_N = \frac{7\mbox{\,AU}}{\tau} \cdot \exp(-t/\tau) \le \frac{7\mbox{\,AU}}{\tau}.
  \label{eq:timescale}
\end{equation}
According to the adiabatic invariant theory, to permit planetesimals to be captured into the 2:3 NMMR, the value of $\dot{a}_N$ should be smaller than $\sim10^{-5}$\,AU/yr for initial eccentricities of $0.05-0.3$ (Melita \& Brunini 2000). Namely, from equation (\ref{eq:timescale}), the migration timescale $\tau$ is demanded to be at least on the order of $10^6$ yr. With this in mind, we choose an order of magnitude larger $\tau=2\times10^7$ yr. This value is derived from the orbital evolution of Jovian planets embedded in a self-gravitating planetesimal disk in our previous work (Li et al. 2011). 

Secondly, we would like to rule out the $a$-dependence. In numerical simulations we put all test particles initially at $a=33$\,AU, where exterior to the initial 2:3 NMMR centered at $\sim30.4$\,AU and interior to the initial 1:2 NMMR at $\sim36.8$\,AU. In this way test particles can avoid the dynamical stirring from the sweeping 1:2 NMMR ahead; thereafter they would stay on approximately original orbits and await the forthcoming 2:3 NMMR. Besides, the strong $\nu_8$ secular resonance initially at 29.9\,AU (Li et al. 2008) could also be avoided. It is reasonable to assume that the planetesimal disk was truncated around 35\,AU (Gomes et al. 2004; Li et al. 2011). As the 2:3 NMMR sweeps from about 30.4 to 35\,AU, its migration speed varies slightly and is always slow enough to be adiabatic. Consequently, the planetesimals throughout this region likely face the similar fate as the ones located at 33\,AU. We think that our equal-$a$ simulations may provide important insights into the orbital evolution of inclined Plutinos. 

Thirdly, with regard to the $e$-dependence, the probability of resonant capture has been given by Dermott el al. (1988). If a planetesimal encounters the 2:3 NMMR having $e$ smaller than a critical value $e_{\mbox{\footnotesize{crit}}}\sim0.052$ (for $\dot{a}_N<2.9\times10^{-5}$\,AU/yr), then the adiabatic resonant trapping is certain. In consideration of $e$ fluctuations created by solar and planetary perturbations, the initial $e$ is taken to be 0.01, ensuring test particles to approach the 2:3 NMMR with $e<e_{\mbox{\footnotesize{crit}}}$.

On the basis of above arguments, we performed a series of runs in which the initial inclination $i_0$ is treated as the only adjustable parameter. For each value of $i_0$ between $10^{\circ}$ and  $90^{\circ}$ ($\Delta i_0=10^{\circ}$) in a run, 200 test particles with the same $a$ (33\,AU) and $e$ (0.01) are introduced. The other three angles $\Omega$, $\varpi$ and $\lambda$ of an orbit are randomly sampled in the range 0--2$\pi$. The duration of the integration spans $5\tau=10^8$ yr; at this time, the four planets have reached their current configurations. In the orbital calculations that include planetary migration here, we also adopt the swift\_rmvs3 integrator with a time step of 0.5 year as in Section 3.

\subsection{Capture into the 2:3 NMMR}

Gomes (1997) indicated that the expression 
\begin{equation}
K=\sqrt{a}\left( \sqrt{1-e^2}\cos{i} - q/p \right)
\label{Kvalue}
\end{equation}
is almost constant after the planetesimal has been trapped into the $q:p$ mean motion resonance. This can be deduced explicitly from the disturbing function or the Jacobi constant in the restricted three-body problem. Thus if the planetesimal falls into the 2:3 NMMR from the nearly-circular and coplanar orbit, and is subsequently pushed outwards by this resonance during Neptune's migration, its $e$ and $i$ would obey the relation
\begin{equation}
\sqrt{1-e^2}\cos{i} = \left( 2 + \sqrt{a_N^{(i)}/a_N^{(f)}} \right)/3,
\label{EIvalue}
\end{equation}
where $a_N^{(i)}$ is the semimajor axis of Neptune at the moment of resonant capture, and $a_N^{(f)}$ is the final value about 30\,AU. By looking at equation (\ref{EIvalue}), if $i$ increases from $0^{\circ}$ to $30^{\circ}$ during the phase of resonance lock, Neptune has to begin its journey interior to 10.7\,AU. This seems quite doubtful in the prevalent understanding of the Solar system formation. As a matter of fact, Gomes (2000) showed that a Plutino with modestly high inclination ($i>15^{\circ}$) cannot be produced by the combination of the 2:3 NMMR, the Kozai mechanism and the $\nu_{18}$ secular resonance. This suggests that the primordial planetesimal disk was dynamically excited before resonance sweeping, thus the initial conditions of high $i_0$ that we adopted is reasonable.

A wealth of information on the 2:3 NMMR sweeping of the pre-inclined planetesimals is given in Table 1. The first column ($i_0$) indicates the initial inclinations of test particles; and these values would barely change until the 2:3 NMMR arrives. For each $i_0$, 200 test particles with the same initial $a$ and $e$ but different angular variables were used to derive the data presented here. The second column ($R_{2:3}$) is the probability of capture into the 2:3 NMMR, only regarding objects with $A_{\sigma}<120^{\circ}$. According to our results in Section 3, these resonant objects (hereafter ``captured Plutinos'' ) are likely to survive over the age of the Solar system, thus they are only of concern to our study. It is possible that some fraction of captured Plutinos may lose stability in the subsequent long-term evolution, virtually that is a rare case in one extended integration up to $4\times10^9$ yr. The third and fourth columns record, among captured Plutinos, the fractions of objects experiencing the Kozai mechanism with $\omega$ librating around  $90^{\circ}$ ($K_{90}$) and $270^{\circ}$ ($K_{270}$). The next three columns give the ranges of final $e$ and $i$, and the minimum resonant amplitude Min$A_{\sigma}$. The last column gives the number of captured Plutinos who underwent a long enough period of the temporary 7:11 inclination-type NMMR\footnote{The critical argument of the 7:11 inclination-type NMMR is $7\lambda_N - 11\lambda + 4\Omega$, and the associated term in the expansion of the disturbing function is of order $i^4$.} ahead and have resultant small $e$ $(<0.1)$; and for convenience the fraction of these objects out of our initial sample of 200 objects at each $i_0$ is also indicated in parentheses.

For population \Rmnum1 ($i_0\le40^{\circ}$) with inclinations in the range of the observed values (see Fig. \ref{observed}), the capture efficiency $R_{2:3}$ fails from $58.5\%$ for $i_0=10^{\circ}$, to $10.5\%$ for $i_0=40^{\circ}$. This seems to be compatible with the unbiased inclination distribution of the Kuiper belt determined by Brown (2001). He found that there is a tendency to have fewer Plutinos concentrated at increasing inclinations for $10^{\circ}\le i \le 40^{\circ}$. It should be noted that $R_{2:3}$ merely reflects, regardless of other dynamical processes, the influence of $i$ on the capture and retainment of the sweeping 2:3 NMMR. While the intrinsic capture efficiency is also affected by the overlap of NMMRs when Neptune's eccentricity was large (Levison et al. 2008), the scattering encounters with Jovian planets or massive planetesimals, the competition of the 1:2 NMMR, etc. However, these factors are beyond the scope of our paper and will not be discussed here.

After capture into the 2:3 NMMR, a subset of population \Rmnum1 is trapped into the temporary Kozai mechanism. The $e$-$i$ oscillation may cause an irreversible $i$ increase if the coupled large changes in $e$ and $i$ are broken before one complete Kozai cycle is accomplished. Our results show that the magnitude of $i$-excitation is limited to less than $5^{\circ}$ at the end of the integration. This is very important since it could constrain any mechanism of the origin of high-inclination Plutinos, providing the requisite pre-heating in $i$. Opposite to the variation of $i$, $e$ suffers damping due to the Kozai effect, but could hardly attain value below 0.1. 

From the evolution of population \Rmnum2 ($i_0\ge50^{\circ}$), we find that $R_{2:3}$ increases from a minimum $9.5\%$ for $i_0=50^{\circ}$, to $31\%$ for $i_0=80^{\circ}$ and $90^{\circ}$. In contrast with the case of population \Rmnum1, the turnover in $R_{2:3}$ at larger $i_0$ is unexpected. One immediate conjecture is that, the variation of $R_{2:3}$ may have some potential link to Min$A_{\sigma}$ (see Table 1), which decreases with increasing $i_0$ for captured Plutinos from population \Rmnum2. Naturally, with more inclined orbits, certain objects possessing relatively smaller $A_{\sigma}$ could be appended to the 2:3 NMMR zone. Similarly, this hypothesis can also be accounted for the variation of $R_{2:3}$ for population \Rmnum1, as Min$A_{\sigma}$ increases with $i_0$. For the inclination up to $30^{\circ}$, the same correlation between $i$ and Min$A_{\sigma}$ has already been predicted in Wiegert et al. (2003) and Lykawka \& Mukai  (2007).

\begin{figure}
 \vspace{-4mm}
 \hspace{-6mm}
  \includegraphics[width=9.5cm]{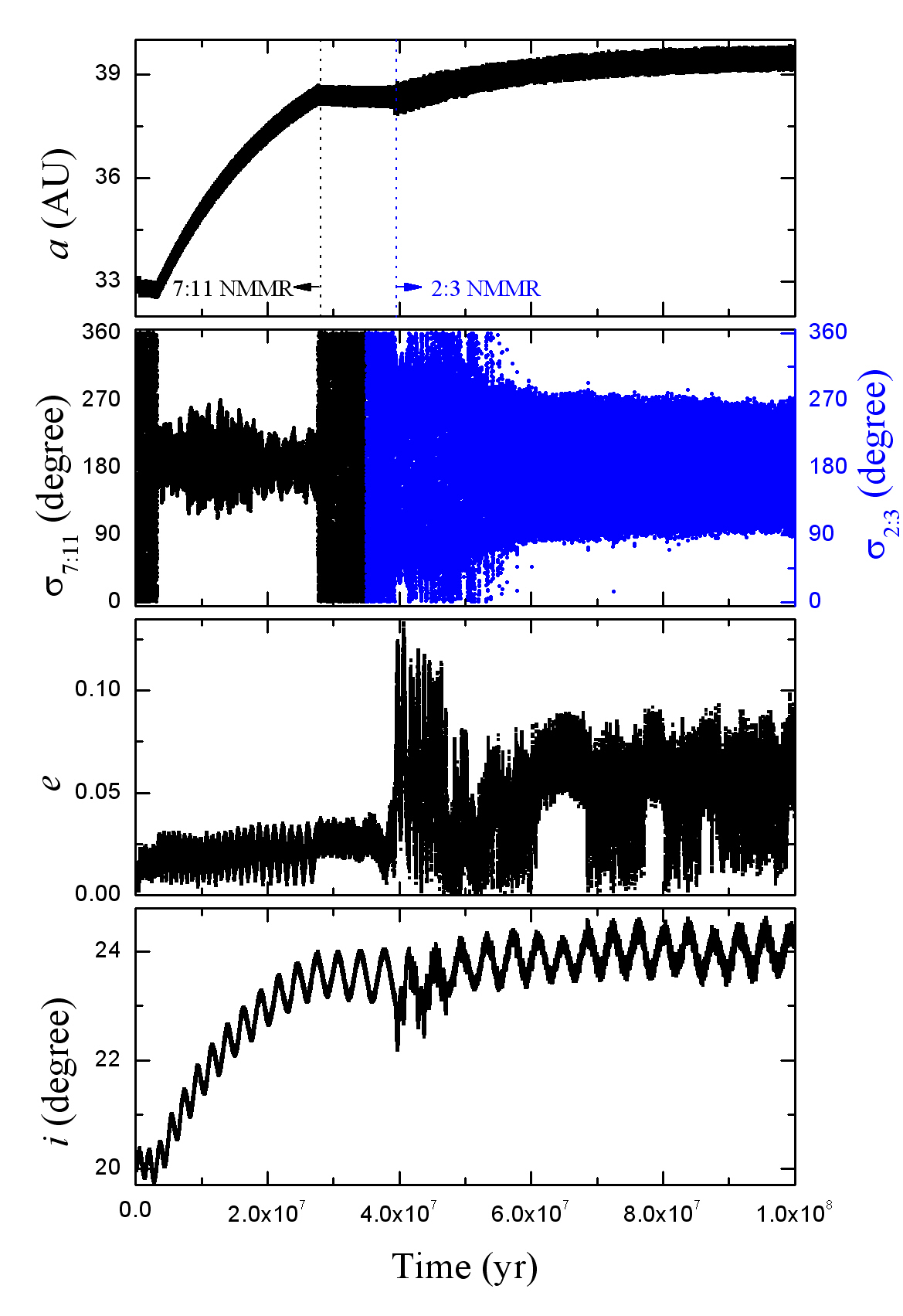}
  \caption{An example of the creation of a low-eccentricity Plutino that starts from an inclined orbit ($i_0=20^{\circ}$). The four panels down from the top show the time evolution of semimajor axis $(a)$, the 7:11 resonant argument ($\sigma_{7:11}=7\lambda_N - 11\lambda + 4\Omega$) together with the 2:3 resonant argument ($\sigma_{2:3}=\sigma=2\lambda_N-3\lambda+\varpi$), the eccentricity $e$ and the inclination $i$.  The planetesimal is first pushed outwards by the 7:11 inclination-type NMMR, and exhibits small eccentricity oscillations during the migration. Near the location around 38.5\,AU, the planetesimal is released from the 7:11 NMMR and then recaptured by the following 2:3 NMMR.} 
  \label{7to11MMR}
\end{figure}

Of captured Plutinos from population \Rmnum2 in our simulations, all have their $i$ decreased by a few degrees. Because these particles are subject to conservative forces, their $K$ values are fixed during the orbital evolution in resonance lock. By making the derivative of equation (\ref{Kvalue}), the time variation of $i$ can be written as 
\begin{equation}
\frac{di}{dt}=\left(\frac{L}{2a}\dot{a} - \frac{e\cos{i}}{\sqrt{1-e^2}}\dot{e}\right)  /  \left(\!\sqrt{1-e^2}\sin{i}\!\right),
\label{doti}
\end{equation} 
where $L=\left(\sqrt{1-e^2}\cos{i} - 2/3\right)$. Since both $a$ and $e$ are driven up by the migrating 2:3 NMMR, we have $\dot{a}, \dot{e}>0$. This implies that the direction of the change of $i$ is determined by the sign of $L$ according to equation (\ref{doti}). For a planetesimal having $\cos{i}<2/3$  (i.e., $i\gtrsim48^{\circ}$),  $L$ is always negative, thus its $i$ must decrease, as in the case of population \Rmnum2. At this point we want to emphasize that, such decrease in $i$ from the very beginning of the 2:3 NMMR trapping is not the Kozai effect, and it continues until the end of the migration.

It is of great interest that, we observe in the simulations some captured Plutinos moving on nearly-circular orbits with $e<0.1$. As we discussed above, the Kozai mechanism alone seems unable to bring Plutinos to such a low-$e$ mode.  A careful checking on the possible dynamical mechanisms shows that the major factor is the 7:11 inclination-type NMMR.  A typical example has been shown in Fig. \ref{7to11MMR}. At about 3 Myr after the start of the integration, the planetesimal is captured into the 7:11 inclination-type NMMR, as diagnosed by the libration of the critical argument $\sigma_{7:11}=7\lambda_N - 11\lambda + 4\Omega$ (second panel from the top). This resonance can only involve the inclinations of trapped objects but not the eccentricities. Thus a low-$e$ state ($e<0.04$) has achieved during the planetesimal's outward migration, while $i$ flips from $20^{\circ}$ to $\sim24^{\circ}$. The 7:11 inclination-type NMMR maintains up to around $2.8\times10^7$ yr, and then the planetesimal diffuses out of this resonance near 38.5\,AU. In the subsequent evolution, the planetesimal stays there until the following 2:3 NMMR picks it up around $4\times10^7$ yr, and it finally reaches a Plutino-type orbit near 39.4\,AU. Despite somewhat excitation by the migrating 2:3 NMMR, the planetesimal's mean eccentricity still undergoes low value episode below 0.1 and ends at roughly 0.055.

The above two-step capture process is very suggestive of the creation of low-eccentricity, high-inclination Plutinos, and the number of such objects ($N_{7:11}$) at different $i_0$ has been given in the last column of Table 1. As for the captured Plutinos with $e<0.1$ generated in our model, all of these objects have experienced the 7:11 NMMR dynamics. However, the resonant capture by the sweeping 7:11 inclination-type NMMR is possible only for highly inclined planetesimals, because the resonance's strength is proportional to $\sin^4(i/2)$ according to the disturbing function (Murray \& Dermott 1999). Consequently, no trapping into this high-order resonance could occur for planetesimals having relatively small $i$ (e.g., the case of $ i_0=10^{\circ}$). In order to better define a threshold of the inclination for triggering this resonance, we have conducted additional runs with  $i_0<20^{\circ}$, under a much higher resolution of $\Delta i_0= 1^{\circ}$. We placed 1000 test particles at each $ i_0$, and evolved the system using the same numerical procedures. The outcome yields that when $i_0\ge 15^{\circ}$, the 7:11 inclination-type NMMR appears, and the particular Plutinos with $e<0.1$ can be generated at the level of $\sim1\%-4\%$ out of a sample of 1000 at different $i_0$. This range of a few percent is consistent with the obtained values in Table 1, i.e.,  $1\%$, $2.5\%$ and $1.5\%$ for $i_0= 20^{\circ}$, $30^{\circ}$ and $40^{\circ}$, respectively.

We further note that for population \Rmnum2, none of captured Plutinos has small final $e$. Although the first step of the resonant capture by the 7:11 inclination-type NMMR could take place, the duration of this event is too short. The planetesimals would be released from the 7:11 inclination-type NMMR not far away from their original locations, and then recaptured by the 2:3 NMMR, in which longer radial shift results in larger $e$.

\subsection{Kozai mechanism}

For most of captured Plutinos coming from our simulations, their $\omega$ librate when they are close to the inner branch of the separatrices of the 2:3 NMMR. This phenomenon can be accounted for their high $i$. Indeed, Wan \& Huang (2007) pointed out that, if an object is inside the 2:3 NMMR under the present orbital configuration of Neptune, the minimum inclination required for the appearance of the Kozai mechanism is just about $10^{\circ}$. In our model the critical inclination should be even smaller as Neptune starts closer to the Sun. However, since the near-separatrix motion is chaotic, the occurrence of the Kozai mechanism at the very early stage of the migration is transient and can only last over several Kozai cycle periods.

As the orbital migration of test particles locked in the 2:3 NMMR continues, a number of them find their way back to the Kozai mechanism. Recorded in Table 1, a subset of captured Plutinos keep on the Kozai mechanism until the end of the integration. The libration center $\omega_0$ groups about two values: $90^{\circ}$ and 270$^{\circ}$. While no captured Plutinos are found to survive with $\omega$ librating around $0^{\circ}$ or $180^{\circ}$, which are unstable equilibrium points as shown in the theoretical analysis by Wan \& Huang (2007). 

Of captured Plutinos from population \Rmnum1, the probability of trapping into the Kozai mechanism in a stable state is significant. Considering the number of captured Plutinos at different $i_0$, a weighted fraction of these $\omega$-librators is estimated to be $\sim56\%$, which is about three times of that of observed samples. Although the fraction of Plutinos experiencing the Kozai dynamics is biased in the observations, Gladman et al. (2012) discuss this bias in detail and point out that the intrinsic up limit should be less than  $33\%$. In fact, we neglect some physical factors in our model, e.g., the stochastic effects in the planet migration (Zhou et al. 2002), the collisions among Plutinos in the primordial evolution (Kenyon et al. 2008), etc., all of which would reduce the capture of planetesimals as Plutinos in general and might be possible to affect the fraction of Kozai Plutinos. Anyway, we have confirmed that the Kozai mechanism associated with the sweeping 2:3 NMMR is efficient for planetesimals with $i_0$ up to $40^{\circ}$.  Moreover, as stated below, the capture of Plutinos into Kozai orbits would be only $\sim21\%$ for $i_0=50^{\circ}$; even so, this is also a significant fraction.

\begin{figure}
 \vspace{-4mm}
 \hspace{-5mm}
  \includegraphics[width=9cm]{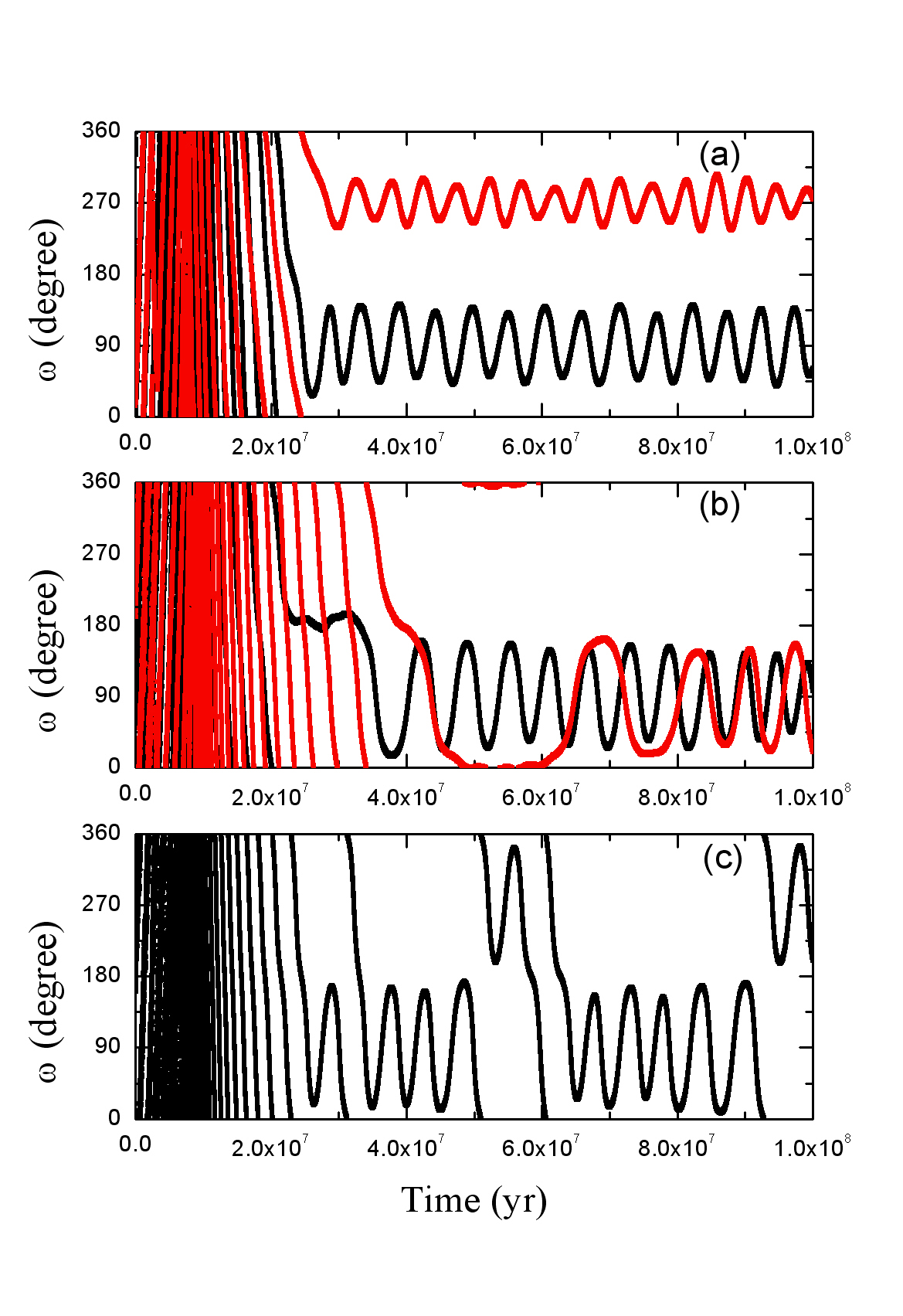}
  \caption{Three typical scenarios for the time evolution of the argument of perihelion ($\omega$) for captured Plutinos associated with the Kozai mechanism. (a) Stable libration of $\omega$ around $90^{\circ}$ or $270^{\circ}$, with amplitudes smaller than $90^{\circ}$. (b) Transformation of the $\omega$-libration center from $0^{\circ}$ or $180^{\circ}$ (unstable equilibrium points) to $90^{\circ}$ or $270^{\circ}$. (c) Frequent alternation between oscillations of $\omega$ at $90^{\circ}$ and $270^{\circ}$. The black and red curves in panels (a) and (b) refer to two distinct objects, respectively.}
  \label{kozai}
\end{figure}

Fig. \ref{kozai} displays three scenarios for the evolution of $\omega$-librators in the 2:3 NMMR, by taking the case of $i_0=20^{\circ}$ for illustration: (1) they exhibit typical motion for the Kozai mechanism around $\omega_0=90^{\circ}$ or $270^{\circ}$, with amplitudes $\Delta\omega<90^{\circ}$ until the end of the integration (Fig. \ref{kozai}a); (2) they may originally experience the Kozai mechanism around the unstable equilibrium points $0^{\circ}$ or 180$^{\circ}$ for about $1-2$ Myr, and finally their $\omega_0$ switch to the stable ones at 90$^{\circ}$ or 270$^{\circ}$ (Fig. \ref{kozai}b); (3) the libration center $\omega_0$ varies from time to time about 90$^{\circ}$ and 270$^{\circ}$, thus $\Delta\omega$ would reach 360$^{\circ}$ during the evolution (Fig. \ref{kozai}c). Based on the last scenario, the circulation of $\omega$ cannot be simply judged by the 360 degrees difference of its maximum and minimum values, and we should be very careful to extract the possible orbits affected by the Kozai dynamics.

Given the results in Table 1, we can see an asymmetry in the distribution of the libration center $\omega_0$ associated with the Kozai mechanism: Plutinos are more likely to have their $\omega$ librating around $90^{\circ}$ than around $270^{\circ}$. However, the difference in the number is at the level of $\lesssim150$\% and it is not too pronounced. Furthermore, the alternation between oscillations of $\omega$ at 90$^{\circ}$ and 270$^{\circ}$ is frequent for Plutinos experiencing the Kozai mechanism. An example of such event has been shown in Fig. \ref{kozai}c, where $\omega_0$ switches to 270$^{\circ}$ at the end of the integration. Actually, this small and time-evolving asymmetry in the $\omega_0$-distribution can also be detected in observed Plutinos, so any constraint on the migration of Neptune may not be provided here. A large enough size of samples should be explored later to draw definite conclusions of this asymmetry. 

As for population \Rmnum2, the capture efficiency of the Kozai mechanism shows a sharp drop to about 21\% at $i_0=50^{\circ}$, and becomes zero at $i_0\ge60^{\circ}$. We find that this is fated to happen. If the trapping into the Kozai mechanism took place, $e$ and $i$ of the particle would start to suffer large opposite oscillations. In every oscillating cycle of $\omega$, $i$ would undergo large increase phase on the timescale of million years, which is in conflict with the fact that a particle with $i\gtrsim48^{\circ}$ locked in the outward-migrating 2:3 NMMR must have its $i$ decreased all the time according to equation (\ref{doti}). Thus the coupling of the 2:3 NMMR and the Kozai mechanism is unlikely to maintain for population \Rmnum2. Unless for a few lucky captured Plutinos with $i_0=50^{\circ}$, once their $i$ decrease to below $48^{\circ}$ forced by the sweeping 2:3 NMMR, they have opportunities to encounter the Kozai mechanism. Here we would like to stress that, the above argument does not exclude a temporary libration of $\omega$ near the separatrix of the 2:3 NMMR (e.g., at the beginning of the resonance encounter mentioned previously), where the motion is chaotic.

A comment on the validity of equation (\ref{doti}) for the cessation of the Kozai mechanism must be made here.  As shown in Gomes (1997), an equivalent form of equation (\ref{doti}) can be derived using the explicit disturbing function, which considers angels
\begin{equation}
\sigma_i=q_i\lambda_N-p_i\lambda+k_i\varpi+l_i\Omega,
\label{explicit}
\end{equation} 
where $q_i/p_i=q/p=2/3$ and $q_i, p_i, k_i, l_i$ are integers. If an object is experiencing both the 2:3 NMMR and the Kozai mechanism, the associated critical arguments can be expressed as (Wan \& Huang 2007)
\begin{equation}
\theta=j_1\sigma+j_2\omega=2j_1\lambda_N-3j_1\lambda+(j_1+j_2)\varpi-j_2\Omega,
\label{23andkozai}
\end{equation} 
where $j_1, j_2$ are integers and $j_1\ge1$, $j_2\ge2$. It is quite obvious that $\theta\in\{\sigma_i\}$, and therefore, equation (\ref{doti}) does include the dynamics of the Kozai mechanism inside the 2:3 NMMR.  So the above result is valid for explaining the absence of Kozai Plutinos with $i\gtrsim48^{\circ}$.


\section{Conclusions and discussion}

In recent years, a few KBOs on extremely inclined orbits ($i>40^{\circ}$) have been discovered. Thus it is of great interest to build up a picture of the possible high-inclination population throughout the Kuiper belt. In this paper, we present the results of the study of the dynamics of Plutinos with $i$ up to $90^{\circ}$.

By making use of a semi-analytical model based on the simplified disturbing function, we first investigated the particular features of the 2:3 NMMR for inclined orbits. We have defined the special libration center (SLC) of the resonant angle $\sigma$ at the minimum of the averaged disturbing function $R(\sigma)$. Unlike low-inclination Plutinos, for high-inclination ones their SLCs are no longer fixed but strongly dependent on $\omega$ (Gallardo 2006b). Nevertheless, there is a general libration center (GLC) $\sigma_0$ at $180^{\circ}$, which is achieved over the timescale several times longer than a single libration period. The maximum excursion of $\sigma$ from $\sigma_0$ measures the resonant amplitude $A_{\sigma}$. We have shown that high-inclination Plutinos always possess substantial $A_{\sigma}$ due to the variation of the SLCs, thus they would never settle at the exact 2:3 NMMR (i.e., $A_{\sigma}=0^{\circ}$). The resonant amplitude's lower limit $\Delta\sigma_{min}$ increases with $i$, and achieves the value about $75^{\circ}$ when $i\sim80-90^{\circ}$.

Under the perturbations of four Jovian planets, the orbits of thousands of test particle Plutinos with high inclinations ($10^{\circ} \le i \le 90^{\circ}$) were integrated for times up to the age of the Solar system. We find that the inclinations of stable orbits inside the 2:3 NMMR could be as high as $90^{\circ}$. In the dynamical map on the initial ($a, i$) plane, the stable regions are mainly delimited by the resonant amplitude $A_{\sigma}$ less than about $120^{\circ}$, and this critical value is always larger than the prescribed lower limits $\Delta\sigma_{min}$. 

We also notice some complex dynamical structures in the 2:3 NMMR. For the case of small $e$, outside the narrow stable bands extending to $i=90^{\circ}$ near the resonance center at $\sim39.4$\,AU, the stable and unstable regions interlace with one another. This is caused by the appearance of the Kozai mechanism, which generally has a destabilizing effect on the high-inclination Plutinos with $e\lesssim 0.1$. While for the case of moderate to large $e$, when $i_1<i<i_2$, the instability occupies the resonance center due to the temporary state of the Kozai mechanism, but a few resonant orbits with relatively large $A_{\sigma}$ could still be stable over 4 Gyr; when $i<i_1$ or $i>i_2$, a wide neighborhood of the resonance center is almost surrounded by stable orbits. The two limits $i_1$ and $i_2$ are only certain for a specific $e$, and both shift to larger values with increasing $e$. Based on our results (see Fig. \ref{stability}), their tentative values are suggested to be $i_1\sim30^{\circ}$ and $i_2\sim50^{\circ}$ for $e=0.15$; $i_1\sim60^{\circ}$ and $i_2\sim80^{\circ}$ for $e=0.3$. These quantities represent boundaries that divide stable and unstable regions within the 2:3 NMMR. 

With respect to the mechanism of the 2:3 NMMR sweeping and capture during Neptune's migration, further simulations have been undertaken to investigate the possibility of the outward transportation of planetesimals on inclined orbits. In our model, Neptune started at 23\,AU with a migration timescale $\tau=2\times10^7$ yr, and all test particles were initially at 33\,AU. The results show that the resonant capture and retainment is allowed for any initial inclination $10^{\circ}\le i_0\le90^{\circ}$. And of surviving objects in the 2:3 NMMR at the end of the migration, we define captured Plutinos as ones moving on resonant orbits with $A_{\sigma}<120^{\circ}$ according to the above stability analysis.

As for captured Plutinos obtained in our simulations, they may be classified into two groups:  population \Rmnum1 with $i_0\le40^{\circ}$ and population \Rmnum2 with $i_0\ge50^{\circ}$. Our principal findings to these two populations can be generalized as follows:

(1) Population \Rmnum1 has the inclination within the range of the observed values for Plutinos. Referred to our migration model, the capture efficiency $R_{2:3}$ of the sweeping 2:3 NMMR decays from $\sim60\%$ for $i_0=10^{\circ}$, to $\sim10\%$ for $i_0=40^{\circ}$. For captured Plutinos with high $i_0$ here, the magnitude of $i$-excitation is less than  $5^{\circ}$ during planetary migration, indicating the existence of a stirred-up transneptunian disk prior to resonance sweeping. Most importantly, some captured Plutinos are found to move on nearly-circular orbits with final $e<0.1$. We find that all of these objects are first captured into the 7:11 inclination-type NMMR. As they migrate outwards locked in this resonance, their $e$ always remain at $\lesssim0.04$, oscillating but undergoing no net growth. When these planetesimals approach the location near $39.4$\,AU, they are released from the 7:11 inclination-type NMMR in the low-$e$ mode, and then picked up by the following 2:3 NMMR. Despite somewhat excitation in the subsequent evolution, their $e$ would end at $< 0.1$. This mechanism for the creation of nearly-circular Plutinos requires $i_0\ge15^{\circ}$.

Among the captured Plutinos from population \Rmnum1, we obtained a total of 10 objects with $e<0.1$ and $i>10^{\circ}$. However, there are only 5 real Plutinos resided in the same subgroup. As a matter of fact, the intrinsic number of such low-$e$ and high-$i$ Plutinos may differ significantly in an unbiased distribution (i.e., bias favors discovery of high-$e$ and low-$i$ objects). Based on the model by Gladman et al. (2012) from the Canada-France Ecliptic Plane Survey (CFEPS), the population of Plutinos with $e<0.1$ is estimated to be about 490 (diameter $>100$ km). Since none of the characterized CFEPS Plutinos with $e<0.1$ has inclination smaller than $10^{\circ}$, the intrinsic number of low-$e$ and high-$i$ ones would be on the order of several hundred and they should draw more attention. Here, we just intend to propose a new mechanism that may generate the low-$e$, high-$i$ Plutinos, while the intrinsic fraction of Plutinos in this particular group remains to be determined by performing more detailed modeling.

(2) With regard to population \Rmnum2, contrary to what one might expect, the capture efficiency $R_{2:3}$ turns to increase with the inclination, from $9.5\%$ for $i_0=50^{\circ}$ to 31\% for $i_0=80^{\circ}$ and $90^{\circ}$. The considerable value of $R_{2:3}$ implies that a much wider $i$-distribution of Plutinos is theoretically possible. Another interesting point with the captured Plutinos from population \Rmnum2 is that, all of them have their $i$ decreased by several degrees. We find that such $i$-damping is certain to occur for any planetesimal with $i\gtrsim48^{\circ}$ caught into the outward sweeping 2:3 NMMR. An important production of this mechanism is to prohibit the trapping of population \Rmnum2 into the coupled 2:3 NMMR and Kozai mechanism, since an inclination increase induced by the latter would not be allowed.

In order to highlight the effects of the high-inclination on the push-out mechanism, we rule out the $a$-dependence in our model, namely setting initial semimajor axis $a_0=33$\,AU for all test particles. In this way the influence of the strong 1:2 NMMR and $\nu_8$ secular resonance could be avoided. However, a discussion of different $a_0$ in the primordial disk is certainly due. Since the migration of Neptune is modeled with exponential behavior, the speed of the 2:3 NMMR sweeping will decay in the same way when moving from its initial location until reaching $\sim39.4$\,AU. As we know, in the standard framework of the resonant capture mechanism during planetary migration, the capture probability of objects into a NMMR depends on the sweeping rate of the resonance. The slower this rate, the higher the capture probability and at the same time the smaller the minimum eccentricity necessary for capture. To assess the role of $a_0$ in the capture and evolution of high-inclination Plutinos, we have made two additional runs with $a_0$ of 31\,AU and 35\,AU. Combining with the case of $a_0=33$\,AU we find that the capture probability $R_{2:3}$ does undergo some degree of change since the 2:3 NMMR sweeps distinct $a_0$ at different rates. In spite of this fact, for each particular $a_0$, the decreasing (increasing) trend of $R_{2:3}$ with the increasing $i_0$ for population \Rmnum1 (\Rmnum2) is still the same as in Table 1; and our other main results reported in Section 4 are hardly affected. It can be expected that, even if a full disk at $30-35$\,AU is considered on further examination, the remarkable influences of the initial high inclination obtained from a single $a_0$=33\,AU are intrinsic. 

In the model tested here for $a_0=33$\,AU, although we managed to avoid the dynamical influence of the important first-order 1:2 NMMR, the same cannot be done about the second-order resonances, i.e., the 3:5 NMMR initial at $\sim32.3$\,AU. At the end of our $10^8$ yr migration simulations, only two of all survived test particles are detected in the 3:5 NMMR.  Both of them are from the case of $i_0=10^{\circ}$, while none of more inclined particles are captured. This suggests that the 3:5 NMMR has a considerably low capture probability for bodies with high inclinations, and plays little role in sweeping the 33\,AU location. However, it is quite possible that other  higher order NMMRs (or even other types of resonances), in addition to the inclination-type 7:11 NMMR, may be important if the full disk is used in the modeling.

With regard to the effect of initial eccentricity $e_0$ on the adiabatic resonance capture into the 2:3 NMMR, we performed one extra run with $e_0=0.05$ for $a_0=33$\,AU. As suspected, even the capture probability $R_{2:3}$ drops off for larger $e_0$, the major conclusions about the $i_0$-dependence of Plutinos that we have found can also be obtained.

In the future work, we are giving a throughout analysis to the high-inclination Twotinos in the 1:2 NMMR, where the resonance structure is much different.


\section*{Acknowledgments}

This work was supported by the Natural Science Foundation of China  (NSFC, Nos. 11003008, 11178006, 11333002), 985 project of Nanjing University and Advanced Discipline Construction Project of Jiangsu Province. LYZ is grateful for the financial support of NSFC under grant No. 11073012  and the National  `973' Project (No. 2013CB834900). SYS has to acknowledge grant provided by the NSFC (No. 11078001) and the National  `973' Project (No. 2013CB834103).  We are grateful to Tabar\'{e} Gallardo for supplying numerical tools for calculating the averaged disturbing function $R(\sigma)$. The authors would also like to express their thanks to the anonymous referee for greatly valuable comments.

\end{document}